\documentclass[twocolumn,showpacs,%
aps,superscriptaddress,
prd,notitlepage,showkeys,
nofootinbib]{revtex4-1}

\usepackage{amssymb}
\usepackage{amsmath}
\usepackage{graphicx}
\usepackage{dcolumn}
\usepackage[colorlinks,urlcolor=blue,citecolor=blue,linkcolor=blue]{hyperref}
\usepackage{color,units}
\usepackage[dvipsnames]{xcolor} 
\usepackage{lineno}
\usepackage{xspace}
\usepackage{longtable} 
\usepackage{float} 

\usepackage{amsfonts,wasysym,epsfig,verbatim,subfigure,bm,mathrsfs,lipsum}

\begin{document}
\newcommand{\IUCAA}{Inter-University Centre for Astronomy and
 Astrophysics, Post Bag 4, Ganeshkhind, Pune 411 007, India}
\title{Tidal heating of Quantum Black Holes and their imprints on gravitational waves}
\author{Sayak Datta}\email{skdatta@iucaa.in} 
\affiliation{\IUCAA}
\date{\today}
\begin{abstract}
The characteristic difference between a black hole and other exotic compact objects (ECOs) is the presence of the horizon. The horizon of a classical black hole acts as a one-way membrane. Due to this nature, any perturbation on the black hole must satisfy ingoing boundary conditions at the horizon. For an ECO either the horizon is replaced or modified with a surface with non zero reflectivity. This results in a modification of the boundary condition of the perturbation around such systems. In this work, we study how tidal heating of an ECO gets modified due to the presence of a reflective surface and what implication it brings for the gravitational wave observations. We argue that the position of the reflective surface, $\varepsilon$ $\gtrsim 10^{-5}$, can have an observational impact in extreme mass ratio inspirals.  We also discuss a possible degeneracy between $\varepsilon$ and reflectivity, $|\mathcal{R}|^2$, in the context of parameter estimation.
\end{abstract}

\maketitle

\section{Introduction}

LIGO's observation of multiple compact binary mergers has initiated the era of gravitational wave (GW) astronomy~\cite{LIGOScientific:2018mvr}. The LIGO-Virgo collaboration has also observed the first binary neutron star merger GW170817~\cite{gw170817}. These observations provided a stimulating boost towards the tests of general relativity in the strong-field regime~\cite{LIGOScientific:2019fpa}. Properties of vacuum spacetime, propagation of GW, violation of Lorentz invariance has been tested rigorously, resulting in stringent bounds on the mass of the graviton and violations of Lorentz invariance \cite{TheLIGOScientific:2016pea, TheLIGOScientific:2016src, Abbott:2017vtc}. It has also become possible to test the nature of the compact objects in an inspiraling binary. The high compactness of these components leads us to the conclusion that they are either black holes (BHs) or neutron stars(NSs). But it has not been proven conclusively if the both of the components are indeed BHs and not some exotic compact objects (ECOs).

To resolve the information-loss paradox Planck scale modifications of black hole horizons and BH structure have been proposed \cite{Lunin:2001jy, Almheiri:2012rt}. Other ECOs i.e. gravastars that have an interior consisting of self-repulsive de sitter spacetime surrounded by an ordinary matter shell, have also been proposed for similar reasons \cite{Mazur:2004fk}. Similarly, there are boson stars, that are ECOs made of scalar fields \cite{Liebling:2012fv}. Therefore it is necessary to understand how to tell them apart from observation. 

To probe the nature of the compact objects in binary, several tests have been proposed. From the post-merger signals, it is possible to distinguish BH and ECOs using {\it echoes} \cite{Cardoso:2016rao, Cardoso:2016oxy, Tsang:2019zra,Abedi:2016hgu, Westerweck:2017hus, Cardoso:2019rvt}. Rigorous modeling and search for {\it echoes} in data has already begun~\cite{Tsang:2019zra,Abedi:2016hgu, Westerweck:2017hus}. Measurement of the tidal deformability~\cite{Cardoso:2017cfl, Sennett:2017etc, Brustein:2020tpg} and the spin induced multipole moments \cite{Krishnendu:2017shb, Datta:2019euh} can also bring a plethora of information that will be useful for this purpose.

In General Relativity, the horizon of the classical BHs ar perfect absorbers \cite{MembraneParadigm,Damour_viscous,Poisson:2009di,Cardoso:2012zn}. This is due to the causal structure of the geometry of BH. This null surface which is the defining feature of a BH is a one-way membrane. Due to the nature of the horizon, the boundary conditions for the perturbations at the horizon are taken to be ingoing boundary conditions \cite{Berti:2009kk}. But in the case of the ECOs, this boundary condition can get modified \cite{Cardoso:2019apo}. This results in the modification of the perturbation quantities, resulting in observable changes. In the current work, we will focus on how these changes will modify the rate of change of mass and the angular momentum of ECOs. 

Change of mass and angular momentum of the ECO will back react on the orbit. This is called tidal heating \cite{Hartle:1973zz,Hughes:2001jr,PoissonWill}. Tidal heating of BH has been studied in several works \cite{Alvi:2001mx, Chatziioannou:2012gq}. In several works, it has been proposed that the tidal heating effects of ECOs will be different from BHs due to the effective reflectivity of the ECOs \cite{Datta:2019euh, Maselli:2017cmm, Datta:2019epe}.

Modification of tidal heating and usage of it for the purpose of distinguishing different kinds of compact objects using both space-based and ground-based detectors has been studied in several works \cite{Datta:2019euh, Maselli:2017cmm, Datta:2019epe, SDLIGO}. These works are based on the assumption that the rate of change of the mass of ECOs are proportional to the change of the mass if it were a BH \cite{Datta:2019euh, Maselli:2017cmm, Datta:2019epe},

\begin{equation}
\label{mass rate assumption}
\dot{M}_{ECO} = (1 - |\mathcal{R}|^2)\dot{M}_{BH},
\end{equation}

where $\mathcal{R}$ is the reflectivity of the ECO and an overdot represents the time derivative. In this work, we focus on studying the validity of this assumption. For this purpose we take the position of the reflective surface at $r = r_s = (1+\varepsilon)r_+$, where $r=r_+=(M+\sqrt{M^2 -a^2})$, is the position of the horizon if it were a black hole (BH). In this work we study how nonzero $\varepsilon$ affects tidal heating, to our knowledge, this has never been addressed before. It is obvious that how the tidal heating effects will be modified that will depend on the specific model of the ECO \cite{Oshita:2019sat}. However, modifying the horizon boundary condition can give a conservative approximation that will help us understand the tidal heating of ECOs better.

In Sec. \ref{framework} we discuss the basic framework and some definitions that are relevant for the paper. In Sec. \ref{area change of KECO} we discuss how the area change of a quantum black hole (QBH) (which is a Kerr like ECO) depends on its physical properties. In Sec. \ref{tidal heating due to stationary companion} we formulate the problem by reviewing Ref. \cite{Alvi:2001mx}. In Sec. \ref{Perturbation of KECO} we discuss the perturbation and it's boundary conditions for a Kerr like ECO (KECO). We also discuss how these modifications will affect the tidal heating of a KECO. In Sec. \ref{energy and angular momentum fluxes `` down the horizon"} we explicitly calculate the rate of change of spin and area of a KECO with a stationary companion. Using the results in Sec. \ref{energy and angular momentum fluxes `` down the horizon"} in Sec. \ref{fluxes down the horizon for KECO in a binary} we calculate the rate of change of area and spin of a KECO in a binary. In Sec. \ref{implication for GW observations} we discuss how the newfound results affect the emitted gravitational wave (GW) of a KECO binary. Finally in Sec. \ref{discussion} we conclude while discussing future prospects.

Throughout the paper, we take $G=c=1$ and the signature is $(+---)$.

\section{Framework}
\label{framework}

In this work I will follow the notations described in the Ref. \cite{Alvi:2001mx}. The 3-vectors will be denoted by boldface letters. A dot between two 3-vectors denotes the inner product in Euclidean 3-space. A hatted 3-vector will be used to represent the unit vector in that direction. In this article, we focus on Kerr-like ECOs (KECOs), QBH is one of such objects. Properties of KECOs will be described in later sections. From now on we will use QBH and KECO interchangeably.

We consider a binary system with the separation $b$ between the components which is much larger than their total mass $M=M_1+M_2$, where $M_i$ represents the mass of the $i$th component. Define $\mu = M_1M_2/M$ and $\eta = \mu/M$. We will label the components as KECO1 and KECO2, and we denote their spins by $\boldsymbol{S}_i$. The magnitude of the spin is $S_i = (\boldsymbol{S}_i.\boldsymbol{S}_i)^{1/2}$. From $S_i$ we define the dimensionless spin parameter $(\chi_i)$ as $S_i = \chi_{i} M_i^2$. A few Newtonian quantities need to be defined: the orbital angular momentum $\bold{L}_N$, the orbital angular velocity $\Omega_N = (M/b^3)^{1/2}$, and the relative velocity $v = (M/b)^{1/2}$.

As the companions are widely separated they have a region surrounding them satisfying,
\begin{itemize}
\item companions are far enough so that the gravity is weak there,
\item the bodies does not extend so far that the companion's tidal field varies appreciably.
\end{itemize}
In such a region it is possible to place a coordinate system in which the component is momentarily are at rest. These coordinates are referred to as the local asymptotic rest frame (LARF) of the component \cite{Thorne:1984mz}. To label the separate regions of the components we will use LARF1 and LARF2.

In general relativity mass and angular momentum of an object is defined globally using the field at infinity. Since we assume that the components are well separated we define their mass and angular momentum in the LARF. For further details check Ref. \cite{Alvi:2001mx}. With the definitions at hand the quantities $dM_i/dt$ and $dS_i/dt$ can be computed from $dA_i/dt$ using the modified version of the first law as described in Sec. \ref{area change of KECO} and the relation $\omega dJ_i = mdM_i$ for Kerr-perturbation modes of angular frequency $\omega$ and azimuthal angular number $m$ \cite{Thorne:1984mz, Teukolsky:1974yv,Hawking:1979ig}. In this case $J_i$ is the angular momentum of the KECO.

In this work, we will focus only on the KECO1. The results for KECO2 can be found by changing the subscripts as $1\leftrightarrow 2$.

\section{area change of KECO}
\label{area change of KECO}

In this work we focus on a ECO model that has Kerr metric with mass $M$ and dimensionless spin $\chi$ outside a certain radius say $r = r_+(1 +\varepsilon)$, where $r_+ = M(1+\sqrt{1-\chi^2})$. Our goal in this paper is to study the tidal heating of ECOs. We assume that due to the modification of the horizon physics, near horizon property changes. 

The area of a BH is calculated at $r=r_+$. The rate of change of the area of a BH, therefore, comes from the evolution of the area of this surface. In the present scenario we have a reflective surface around the black hole at $r = r_s = r_+ (1+ \varepsilon)$. Interesting discussion regarding the reflectivity and the position of the reflective surface can be found here~\cite{DAmico:2019dnn, Addazi:2019bjz}. Intersection of this ``reflective horizon" with $\mathcal{V} = \rm{const.}$ surfaces will be the relevant two surfaces of a KECO, where $\mathcal{V}$ is the advanced time coordinate. From now on the area of this reflective surface will be considered as the area of the KECO. Therefore, the induced metric on the $2-$surface becomes,

\begin{equation}
    -\Bar{g}_{AB} d\theta^A d\theta^B= \frac{\Sigma}{\rho^2} \sin^2\theta d\phi^2 + \rho^2 d\theta^2,
\end{equation}

whre, $\Sigma = (r_s^2+a^2)^2 - a^2\Delta \sin^2\theta$, $\rho^2 = r_s^2 +a^2 \cos^2\theta$, and $\Delta = r_s^2 +a^2 -2Mr_s$.

Using the induced metric on this surface the area can be calculated. The area is as follows,

\begin{equation}
    A = \oint_{\mathfrak{R}}\Bar{g}^{1/2} d\theta d\phi = \oint_{\mathfrak{R}}\Sigma^{1/2} \sin \theta d\theta d\phi,
\end{equation}

where $\mathfrak{R}$ is the two dimensional cross section of the reflective surface, described by $\mathcal{V} = \rm{const.}$, $r = r_s$, $0\leq \theta \leq \pi$, and $0\leq \phi <2\pi$.

\begin{equation}
    A = \frac{\pi}{a\sqrt{\Delta}} [2 \sqrt{\Bar{\rho}} a\sqrt{\Delta}-(\Bar{\rho}-a^2\Delta) \log \left(\frac{\sqrt{\Bar{\rho}}-a\sqrt{\Delta}}{\sqrt{\Bar{\rho}}+a\sqrt{\Delta}}\right)],
\end{equation}

where $\Bar{\rho} = (r_s^2 +a^2)^2$.

Owing to the smallness of $\varepsilon$ it is possible to expand all the $r$ dependencies in the expression of the area in powers of $\varepsilon$. This will give an expression of $A$ in the power series of $\varepsilon$. The result is as follows:

\begin{equation}
A = \sum_{i=0}^{\infty} \varepsilon^i A^{(i)},
\end{equation}

\begin{eqnarray}
A^{(0)} =& 8\pi M r_+,\\
A^{(1)} =& \frac{4\pi r_+}{3M}[r_+^2 + 2M^2 +3Mr_+],\\
A^{(2)} =& \frac{2\pi r_+}{15M^3}[a^2r_+^2 + 6Ma^2r_+ + 10M^2r_+^2\\
&+12M^3r_+ -4M^4],
\end{eqnarray}where $a_i=M_i\chi_i$. One point should be stressed that this is an approximation of the area in the limit that $\varepsilon$ is small.
The interesting thing to notice is, these results are not too different from the results of a BH. In the limit $\varepsilon \rightarrow 0$ this reproduces the area of a BH. Like BH these results are also simple. As expected they depend only on the mass, spin, and $\varepsilon$. Therefore this can be considered as the effect of the modified version of the no-hair theorem where the modification arises due to the $\varepsilon$. 

From the expression of the area, it is straightforward to calculate the area change. The area change of KECO $(\delta A)$ can be expressed as,
\begin{eqnarray}
\delta A =& \partial_M A \delta M + \partial_a A \delta a, \\
\partial_M A =& \sum_{i=0}^{\infty} \varepsilon^i\partial_M A^{(i)},\\
\partial_a A =& \sum_{i=0}^{\infty} \varepsilon^i\partial_a A^{(i)},
\end{eqnarray}
where $(i)$ represents ith order term in the series. $\delta M$ and $\delta a$ is the change in mass $M$ and angular momentum respectively. The first few terms can be expressed as,

\begin{eqnarray}
\partial_M A^{(0)} =& \frac{8\pi r_+^2}{\sqrt{M^2-a^2}}, \\
\partial_M A^{(1)} =& \frac{4\pi r_+^2}{3M^2\sqrt{M^2-a^2}}[2Mr_+ + a^2 + 8M^2].
\end{eqnarray}

\begin{eqnarray}
\partial_a A^{(0)} =& \frac{-8\pi Ma}{\sqrt{M^2-a^2}}, \\
\partial_a A^{(1)} =& -\frac{4\pi a}{3M \sqrt{M^2-a^2}}[2M^2 + 6Mr_+ + 3r_+^2].
\end{eqnarray}

This result is almost similar to that of a BH. The only difference is the coefficients of $\delta M$ and $\delta a$ depends on $\varepsilon$ perturbatively. These results will be used in the later sections to calculate the rate of change of mass and spin of the KECOs.

\section{tidal heating due to stationary companion}
\label{tidal heating due to stationary companion}

In this section, we will discuss the tidal distortion of KECO1 when KECO2 is held stationary. This is almost similar to the calculations done in Ref. \cite{Alvi:2001mx}. Therefore this section can be considered as the review of the calculations done in Ref. \cite{Alvi:2001mx}. Calculation of the tidal distortion involves solving for the Weyl tensor $\psi_0$, using Teukolsky formalism \cite{Teukolsky:1973ha}. With the $\psi_0$ at hand rates of change KECO1 parameters are calculable in a similar way as described in Ref. \cite{Hawking:1972hy,Teukolsky:1974yv}.  
First, we calculate KECO2's tidal field as seen in LARF1 (Local asymptotic rest frame of the companion 1). For this purpose, we will consider only the lowest order Newtonian tidal field that is constant in the LARF1. Take a Euclidean 3-space with a stationary body with mass $M_2$ at coordinate location $(b,\theta_0, \phi_0)$ in a spherical coordinate system. The Newtonian gravitational field in such coordinate can be expressed as,

\begin{equation}
\begin{split}
\Phi(r,\theta,\phi) = -4\pi \frac{M_2}{b}\sum_{l=0}^{\infty}\sum_{m=-l}^{l} &(2l +1)^{-1} (\frac{r}{b})^{l} Y^*_{lm}(\theta_0,\phi_0)\\
&\times Y_{lm}(\theta, \phi),
\end{split}
\end{equation}

for $r<b$.
As we will evaluate the body's tidal field near the origin $r\ll b$. We will focus only on $l=2$ part of the field.  In the Cartesian coordinate the tidal field can be expressed as, $\mathcal{E}_{ij} = \partial_i\partial_j\Phi^{(l=2)}$. After the derivatives are taken it is straight forward to calculate the components in spherical orthonormal coordinates. The combination that is relevant for our purpose is as follows \cite{Alvi:2001mx}
\begin{equation}\label{tidal field combination}
\mathcal{E}_{\hat{\phi}\hat{\phi}} - \mathcal{E}_{\hat{\theta}\hat{\theta}} -2i\mathcal{E}_{\hat{\theta}\hat{\phi}} = 8\pi \sqrt{\frac{6M_2}{5b^3}}\sum_{m=-2}^{m=2}\, _2Y_{2m}(\theta, \phi) Y^*_{2m}(\theta_0,\phi_0),
\end{equation}
where $_2Y_{2m}(\theta, \phi)$ spin weighted spherical harmonics \cite{Goldberg:1966uu}.

Now returning to the region near KECO1, including LARF1, we notice that the space-time there can be described as a perturbed Kerr black hole (as long as we are in outside of the reflective surface). Therefore we cover this region with a Boyer-Lindquist chart $(t,r,\theta,\phi)$. We need to solve Teukolsky equation \cite{Teukolsky:1973ha} in this region for $\psi_0$. As for unperturbed KECO $\psi_0$ vanishes asymptotically $(\psi_0$ as $r/M_1 \rightarrow \infty)$, it  would be the combination $\mathcal{E}_{\hat{\phi}\hat{\phi}} - \mathcal{E}_{\hat{\theta}\hat{\theta}} -2i\mathcal{E}_{\hat{\theta}\hat{\phi}}$ of the external tidal field \cite{Alvi:2001mx} for a perturbed KECO \cite{Thorne:1986iy}. Therefore, in our case $\psi_0$ takes this asymptotic form for $M_1 \ll r \ll b$ in LARF1, given the tidal field $\mathcal{E}_{ij}$ is due to the companion.The angular dependence of $\psi_0$ in the LARF1 will be like the one shown in Eq.~(\ref{tidal field combination}) with $\theta$ and $\phi$ as the Boyer-Lindquist coordinate and $\theta_0, \phi_0$ representing the companion's angular coordinates as seen in LARF1. Therefor the boundary condition would be \cite{Alvi:2001mx},

\begin{equation}
\psi_0 \rightarrow \frac{8\pi \sqrt{6}M_2}{5b^3}\sum_{m=-2}^{2}\, _2Y_{2m} (\theta, \phi) Y^*_{2m}(\theta_0, \phi_0)
\end{equation}

for $M_1 \ll r \ll b$. The only thing that remains now is to solve for $\psi_0$ with a proper boundary condition at the reflective surface. We can express $\psi_0$ as,
\begin{equation}\label{separation pf psi0}
\psi_0 = \sum_{m=-2}^2\,  _2Y_{2m}(\theta,\phi)R_m(r),
\end{equation}
subject to appropriate boundary condition for $R_m(r)$ at the reflective surface, that will be described in the next section.

\section{Perturbation of KECO}
\label{Perturbation of KECO}

As discussed in the previous sections we will assume that the surface $r = r_s = r_+(1 +\varepsilon)$ has a non-zero reflectivity. We will consider this as the boundary of the KECO. Therefore, unlike BH we will put a ``mixed boundary condition" comprising of both ingoing and outgoing mode at this surface. As our goal is to calculate the rate of change of the area of the KECO, the relevant quantity for this purpose is the Weyl scalar $\psi_0$ (see appendix \ref{NP appendix}). The governing equation for $\psi_0$ is the Teukolsky equation \cite{Teukolsky:1973ha}.  The equation has two linearly independent solutions, namely $\psi_0^{in}, \psi_0^{out}$. Given a reflective boundary condition, the general solution near $r_s$,
\begin{equation}
\psi_0(r\sim r_s) \sim  {\mathfrak T} \psi_0^{in} + {\cal R} \psi_0^{out},
\end{equation}

where $\psi_0^{in}$ and $\psi_0^{out}$ are respectively the ingoing and outgoing modes and ${\mathfrak T}$ and ${\cal R}$ are the absorption coefficient and the reflectivity of the body. For a BH ${\mathfrak T} \rightarrow 1$ and ${\cal R} \rightarrow 0$.

Under a time-dependent perturbation, the solution for $\psi_0$ in the external region of the reflective surface can be expressed in the following form after redefining $\psi_0^{in}$ and $\psi_0^{out}$ in terms of $Y_{hole}^{in}$ and $Y_{hole}^{out}$ defined in Ref. \cite{Teukolsky:1974yv},

\begin{equation}
\begin{split}
\psi_0 =& {\mathfrak T}\psi_0^{in} + {\cal R} \psi_0^{out}\\
=& \int d\omega \sum_{\ell}\sum_m e^{(-i\omega t +im\phi)}\,\, _2S_{lm}(\theta)({\mathfrak T}Y^{in}_{hole}\Delta^{-2}e^{-ikr^*}\\
&+ {\cal R}Y^{out}_{hole}e^{ikr^*}),
\end{split}
\end{equation}
where $_2S_{lm}(\theta)$ is the $\theta$ dependent part of the spin weighted spheroidal harmonics and $\Delta = r^2 + a^2 -2Mr$.  The relevant quantity for our purpose is the $\psi_0^{HH}$ defined as follows:

\begin{equation}
\begin{split}
\psi_0^{HH} \equiv& \frac{\Delta^2\psi_0}{4(r^2+a^2)^2}\\
=& \int d\omega \sum_{\ell}\sum_m\frac{e^{(-i\omega t +im\phi)e^{-ikr^*}} \,\,_2S_{lm}(\theta)}{4(r^2+a^2)^2}({\mathfrak T} Y^{in}_{hole}\\
&+{\cal R}\Delta^{2} Y^{out}_{hole}e^{2ikr^*}).
\end{split}
\end{equation}

The primary ingredient that is needed to calculate the area change is $\sigma$ \cite{Newman:1961qr, Hawking:1972hy, Teukolsky:1974yv}. In Hawking-Hartle tetrad (HH) (check appendix \ref{Hartle-Hawking tetrad}) $\sigma$ satisfies, (for details check \cite{Newman:1961qr, Teukolsky:1974yv}), 
\begin{equation}
\label{NP sigma eqn main text}
D\sigma^{HH} = 2\epsilon\sigma^{HH} +\psi_0^{HH},
\end{equation}

where $\sigma$ and $\epsilon$ are spin coefficients, described in appendix \ref{NP appendix}.
In the case of KECO due to the $\psi_0^{out}$, there will be an extra contribution to the expression of $\sigma$. This will result in the following modification,

\begin{equation}
\begin{split}
\sigma^{HH} = (D-2\epsilon)^{-1}\psi_0^{HH} &= (D-2\epsilon)^{-1}({\mathfrak T}\psi_0^{in, HH} + {\cal R} \psi_0^{out, HH}).
\end{split}
\end{equation}

Separating them in $\omega, \ell, m$ modes we find,

\begin{equation}
\label{sigma separation}
\begin{split}
\sigma^{HH}_{\omega, \ell, m} = -\frac{{\mathfrak T}\psi_0{}_{\omega, \ell, m}^{in, HH}}{ik+2\epsilon} + \frac{{\cal R}\psi_0{}_{\omega, \ell, m}^{out, HH}}{ik -2\epsilon},
\end{split}
\end{equation}

where $k = \omega -ma/2Mr_+$.

Hawking and Hartle \cite{Hawking:1972hy} showed that for a classical BH,
\begin{equation}
\frac{d^2 A}{dtd\Omega} = \frac{2Mr_+}{\epsilon} |\sigma^{HH}|^2\big|_{r= r_+},
\end{equation}

where $d\Omega$ represents the angular volume. Since $\varepsilon \ll 1$, for KECO approximately we can write,
\begin{equation}
\label{area rate of QBH}
\frac{d^2 A}{dtd\Omega} = \frac{\Bar{g}^{1/2}}{\epsilon_s} |\sigma^{HH}|^2\big|_{r= r_s = r_+(1+\varepsilon)} 
\end{equation}
where $\Bar{g}$ is the determinant of the induced metric on the two sphere, $\Bar{g} = (r_s^2 + a^2)^2 + a^2 \Delta_s \sin^2\theta$ and $\epsilon_s$ is the expression for $\epsilon$ evaluated at $r_s$, 
\begin{equation}
\epsilon_s = M\frac{(r_s^2 - a^2)}{2(a^2+r_s^2)^2}.
\end{equation}

Due to the Eq.~(\ref{area rate of QBH}) area of a KECO will change under a perturbation. We will use this equation to calculate the rate of change of the area of a KECO in the later sections.

\section{energy and angular momentum fluxes `` down the horizon"}
\label{energy and angular momentum fluxes `` down the horizon"}

In the last section, we have prepared the stage for the calculation of the rate of change of the KECO parameters. The difference between a Kerr BH and a KECO is the presence of the reflective surface at $r = r_+(1+\varepsilon)$. As has been discussed in \ref{framework} first we will focus on stationary perturbation $\omega = 0$, and using it we will find our final results. The linearly independent solutions of the Teukolsky equation in the limit of $\omega = 0$ has been derived by Teukolsky (see Eq. (5.7) and Eq. (5.8) in Ch. VI of \cite{Teukolsky_thesis}.
As the boundary condition at the reflective surface has changed the solution of the perturbation can now be written as follows \cite{Teukolsky_thesis} (see appendix \ref{apndx:Teukolsky equation}):

\begin{widetext}
\begin{equation}
\label{R_m}
\begin{split}
    R_m(r) =& C_m\big\{{\mathfrak T} x^{\gamma_m-2}(1+x)^{-\gamma_m-2} F(-4,1,-1+2\gamma_m,-x) + {\cal R} x^{-{\gamma}_m}(1+x)^{\gamma_m} F(0,5,3-2\gamma_m, -x) \big\},\\
    =& C_m\big\{{\mathfrak T} y_1 + {\cal R} y_2 \big\},
\end{split}
\end{equation}
\end{widetext}

where,
\begin{equation}
\gamma_m = \frac{im\chi_1}{2(1-\chi_1^2)^{1/2}}, \,\,\,\,\,x=\frac{r-r_{+_1}}{2M_1(1-\chi_1^2)^{1/2}}
\end{equation}
and $F$ is the hypergeometric function. $y_1$ and $y_2$ are the radial part of $\psi^{in}_0$ and $\psi^{out}_0$.

For classical BH $R=0,\, {\mathfrak T}=1$, therefore we can identify $C_m$ with the result found for BH case in Ref. \cite{Alvi:2001mx}, 

\begin{equation}
C_m = \frac{8\pi M_2}{5b^3\sqrt{6}}\gamma_m(\gamma_m+1)(4\gamma_m^2 -1)Y^*_{2m}(\theta_0,\phi_0).
\end{equation}

Since the second term in the Eq.(\ref{R_m}) is of $\mathcal{O}(\mathcal{R})\mathcal{O}(\varepsilon)$, the dominant contribution will come from the first term in the Eq.(\ref{R_m}). In this paper, we will focus only on the dominant contributions. For this reason, all the results found in this paper are independent of the second term.

Using Eq.(\ref{separation pf psi0}), Eq. (\ref{sigma separation}) Eq.(\ref{area rate of QBH}), Eq. (\ref{R_m}) and $\omega dJ_i = mdM_i$ we find $dM_1/dt = 0$ \cite{Teukolsky:1974yv, Alvi:2001mx} and
\begin{equation}
\label{stationary area rate}
\frac{dA_1}{dt} = \mathfrak{T}^2\sum_{i=0}^{\infty} \varepsilon^i \dot{A}^{(i)}_{\theta_0},
\end{equation}

\begin{equation}
\label{stationary s rate}
\frac{dS_1}{dt} = \mathfrak{T}^2\sum_{i=0}^{\infty} \varepsilon^i \dot{S}^{(i)}_{\theta_0}.
\end{equation}

The detailed expressions can be found in Appendix \ref{Coefficients}.

\section{fluxes down the ``horizon" for KECO in a binary}
\label{fluxes down the horizon for KECO in a binary}

In the previous sections, we have described how tidal heating gets modified due to the presence of a reflective surface. Energy flux down the reflective surface becomes different from the case of a black hole. This result depends not only on the mass and the spin of the KECO but also on the position of the reflective surface $\varepsilon$. In this section, we will discuss how does the energy flux down the surface gets modified when the KECOs are in an inspiraling binary. 

In case of rigid $\phi$ rotation for BH binary formulas for the rate of change of mass and spin of the black hole in terms of horizon integral $I$ is given in Eqs. (7.21) of \cite{Thorne:1986iy}. These formulas have been used to calculate the rate of change of mass and spin in the Ref. \cite{Alvi:2001mx}. An important point to note that the explicit integration of $I$ is not required. The only thing needed is to identify the stationary part of the integral. This point is discussed in detail in the appendix \ref{horizon integral}. In terms of $I$ the results can be expressed in the following form,

\begin{eqnarray}
\label{rates with omega}
\frac{dS_1}{dt} &= (\Omega - \Omega_{H1})I.\\
\frac{dM_1}{dt} &= \Omega  \frac{dS_1}{dt},
\end{eqnarray}

where $\Omega_{H} = \chi/(2r_+)$. An expansion of $I$ in powers of $M_1\Omega$ is of the order of  $v^3$, hence is much smaller then 1. Hence the zeroth order part $I_{0} = I|_{\Omega = 0}$ is independent of $\Omega$ and in our case of binary, can be obtained from the calculations for a stationary companion. From Eq.~(\ref{rates with omega}) we have $\dot{S}_1|_{\Omega = 0} = -\Omega_{H1}I_0$, with overdot representing the time derivative. This can be identified with the expression for $\dot{S}_1$ in Eq.~(\ref{stationary s rate}). Therefore we find,

\begin{equation}
I_0(\theta_0) = \mathfrak{T}^2\sum_{i=0}^{\infty}{\cal I}_0^{(i)}\varepsilon^i.
\end{equation}

Since $I_0$ is the leading order contribution, we will approximate $I$ by the leading order contribution $I_0$ in the paper, along the line of Ref. \cite{Alvi:2001mx}. Assuming the radiation reaction time scale to be long and putting $I_0(\pi/2)$ and $\Omega = (\hat{L}_N.\hat{S}_1)\Omega_N$ in Eq.~(\ref{rates with omega}) we find,

\begin{equation}
\frac{dS_1}{dt} = (\Omega - \Omega_{H1})I_0(\pi/2)=\bigg(\frac{dJ}{dt}\bigg)_N \mathfrak{T}^2\sum_{i=0}^{\infty}{\cal S}^{(i)}\varepsilon^i,
\end{equation}

\begin{equation}
\label{M_1 rate}
\begin{split}
    \frac{dM_1}{dt} = \Omega \frac{dS_1}{dt} =&\bigg(\frac{dE}{dt}\bigg)_N \mathfrak{T}^2 \sum_{i=0}^{\infty}{\cal M}^{(i)}\varepsilon^i\\
    =&\bigg(\frac{dE}{dt}\bigg)_N \mathfrak{T}^2 \sum_{i=0}^{\infty}({\cal M}_5^{(i)}v^5+{\cal M}_8^{(i)}v^8)\varepsilon^i,
\end{split}
\end{equation}
 where, $\mathcal{M}^{(i)}$ represents $i$th order term in the expansion w.r.t. $\varepsilon$. Since for our later purposes we will need post Newtonian (pn) expansion, we expand $\mathcal{M}^{(i)}$ in a series in $v$, where $v$ is the velocity parameter of the pn expansion. ${\cal M}_5^{(i)}$ and ${\cal M}_8^{(i)}$ are respectively the $2.5$pn and $4$pn terms. 
 
\begin{equation}
\label{flux at infinity}
\bigg(\frac{dE}{dt}\bigg)_N = \frac{32}{5}\eta^2 v^{10},\,\,\,\,\,\bigg(\frac{dJ}{dt}\bigg)_N = \frac{32}{5}\eta^2 M v^7,
\end{equation}

and $\eta = M_1M_2/M^2$.

Detailed expressions of the coefficients have been shown in Appendix \ref{Coefficients}.

\section{implication for GW observations}
\label{implication for GW observations}
\subsection{Phasing}

In the last section, we showed how the contribution of tidal heating of KECOs affects the energy loss from the orbit of an inspiraling KECO binary. In this section, we will compute the modification of the phase of the GW emitted by such a system. 

Under the adiabatic approximation, a PN expansion is possible. The dynamics of the system is governed by energy and angular momentum loss from the orbiting system. These dynamics have a contribution considering the components as point particles (PP) and another contribution is due to the finite size effects. The finite-size effects can be decomposed into two main ingredients (i) tidal deformation of an individual component due to the gravitational field of the other component and (ii) the amount of energy absorbed by the individual component from orbit due to tidal heating. The dynamics of the system and therefore the emitted GW depend on all these contributions. Hence, the Fourier transformed GW waveform can be written as follows:
\begin{equation}
\Tilde{h}(f) = \Tilde{A}(f) e^{i(\Psi_{PP}+\Psi_{TD}+\Psi_{TH})}\,,
\end{equation}
where $f$ is the frequency of the GW. $\Tilde{A}(f)$ is the frequency-dependent amplitude of the GW. The phase terms $\Psi_{PP}, \Psi_{TD}$, and $\Psi_{TH}$ are the contributions to the total phase arising from the point-particle approximation, the tidal deformability, and the tidal heating, respectively.

We calculate the phase by using Eq.~(2.7) of Ref.~\cite{Tichy:1999pv}. We found the phase shift due to tidal heating to be as follows,

\begin{equation}
\Psi_{TH} = \frac{3  }{128  \eta v^5} \mathfrak{T}^2 \sum_{i=0}^{\infty} \varepsilon^i \psi^{(i)}.
\end{equation}

The form of the $\psi^{(i)}$ has been shown in Appendix \ref{Coefficients}. For BH the effect of TH (i.e. $\psi^{(0)}$) arises at $2.5$PN order. The contribution due to $\varepsilon$ is also in the similar order as it can be seen from the expression of $\psi^{(1)}$ in Eq. (\ref{psi0}) and Eq.(\ref{psi1}).

This result shows that up to the first power of $\varepsilon$, dependence of phase on reflectivity goes as $1-|\mathcal R|^2$ (assuming that $|\mathfrak T|^2 = 1-|\mathcal R|^2$), as has been assumed in Ref.\cite{Datta:2019epe}. But interestingly, the phasing depends explicitly on the position of the reflective surface $\varepsilon$. As a result, with a sensitive detector, it will be possible to measure the $\varepsilon$ from GW observations. The properties of the ECO will determine the $\varepsilon$. Hence, if both of the ECOs in the binary is of a similar kind then both should have the same value of $\varepsilon$. But note that, even though the dependence on reflectivity is like $1-|\mathcal R|^2$, Eq.~(\ref{mass rate assumption}) is not true beyond $\mathcal O (\varepsilon^0)$.

\begin{figure}
\includegraphics[width=8.0cm]{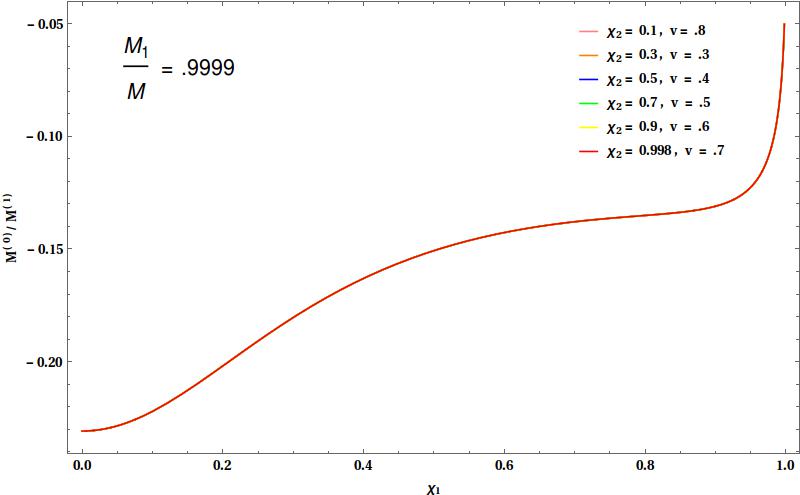}
\caption{${\rm M^{(0)}}/{\rm M^{(1)}}$ has been plotted w.r.t the spins, masses and the post-Newtonian velocity parameter $v$ of the system. The mass ratio is $M_1/M = 9999$
}
\label{EMRI}
\end{figure}

\subsection{Observables: $|\mathcal{R}|^2$ and $\varepsilon$}
In this section, we will focus on a crucial point regarding observability. In several works \cite{Datta:2019epe, Datta:2019euh, Datta:2020gem},  the effect of the reflectivity of the KECO has been considered while ignoring $\varepsilon$. Since the $\varepsilon$ is expected to be very small its contribution regarding tidal heating has been expected to be very small. But the situation can be much more complex than that. Exploiting the smallness of the $\varepsilon$ we have shown that all the relevant physical quantities can be expressed in a perturbative expansion in the power of $\varepsilon$. Therefore, it is quite natural to expect that the contribution of  $\mathcal{O}(\varepsilon) << \mathcal{O}(\varepsilon^0)$. But this is not correct. All the physical non black hole contributions (i.e. $\mathcal{R} \neq 0$ and $\varepsilon \neq 0$) are proportional to $\mathfrak{T}^2$. Assuming that $\mathfrak{T}^2 = 1 - |\mathcal{R}|^2$, all the relevant quantities up to $\mathcal{O}(\varepsilon)$ decomposes into the following structure:

\begin{equation}
\mathcal{O}(\mathcal{R}^0\varepsilon^0) + \mathcal{O}(\varepsilon^0|\mathcal{R}|^2) + \mathcal{O}(\varepsilon\mathcal{R}^0) + \mathcal{O}(\varepsilon|\mathcal{R}|^2),
\end{equation}

where $\mathcal{O}(\mathcal{R}^0\varepsilon^0)$ is the black hole contribution. This implies that during parameter estimation there will be degeneracy between $\varepsilon$ and $|\mathcal{R}|^2$.  This can be emphasized by assuming a system that has $|\mathcal{R}|^2 < \varepsilon \ll 1$. In that case, due to the smallness we can ignore $\mathcal{O}(\varepsilon|\mathcal{R}|^2)$. But $\mathcal{O}(\varepsilon^0|\mathcal{R}|^2)$, $\mathcal{O}(\varepsilon\mathcal{R}^0)$ both are in first order of smallness, resulting in a competitive contribution. Hence it is possible to have a measurable effect due to nonzero $\varepsilon$ while very small reflectivity becomes impossible to measure. In the rest of the section, we will compare these two competing effects and comment on its implications. By the symbol $\mathcal{O}$ we are just representing terms of corresponding powers. I.e. $\mathcal{O}(\varepsilon^0|\mathcal{R}|^2)$ represents the term that is proportional to $\varepsilon^0|\mathcal{R}|^2$. In that sense $\mathcal{O}(\varepsilon^0|\mathcal{R}|^2)$ represents $\mathcal{M}^{(0)}|\mathcal{R}|^2$. $|\mathcal{R}|^2$ has been taken to be small only for the above argument, in general, $|\mathcal{R}|^2$ has not been taken to be small in this paper. 

From all the expressions found in this paper, especially the expression for $\Psi_{TH}$, it can be observed that it is enough to compare between the $\mathcal{M}^{(0)}$ and $\mathcal{M}^{(1)}$. To illustrate it even further,
\begin{equation}
    \begin{split}
        \frac{dM_1}{dt} \propto (\mathcal{M}^{(0)} -|\mathcal{R}|^2 \mathcal{M}^{(0)} + \mathcal{M}^{(1)}\varepsilon - |\mathcal{R}|^2 \mathcal{M}^{(1)}\varepsilon).
    \end{split}
\end{equation}
If we want to understand the importance of $\varepsilon$, then we need to compare only $\mathcal{M}^{(0)}$ and $\mathcal{M}^{(1)}$. But as the systems under consideration are inspiraling binary, it is better to compare the sum of $\mathcal{M}^{(1)}$ of both bodies with the sum of $\mathcal{M}^{(0)}$ of both bodies. So we will compare ${\rm M}^{(1)} \equiv (\mathcal{M}^{(1)}_{\rm body 1} + \mathcal{M}^{(1)}_{\rm body 2})/v^5$ with ${\rm M}^{(0)} \equiv (\mathcal{M}^{(0)}_{\rm body 1} + \mathcal{M}^{(0)}_{\rm body 2})/v^5$.

In Fig. \ref{EMRI}, ${\rm M^{(0)} / M^{(1)}} $ has been plotted for extreme mass ratio inspiral (EMRI). Here the mass of the more massive body has been such that $M_1/M = .9999$, therefore the secondary body is just a point particle. From the plots it is clear that ${\rm |M_0| < | M_1|} $. The consequence of this will be discussed later. 

\begin{widetext}
\begin{figure*}
\includegraphics[width=7.35cm]{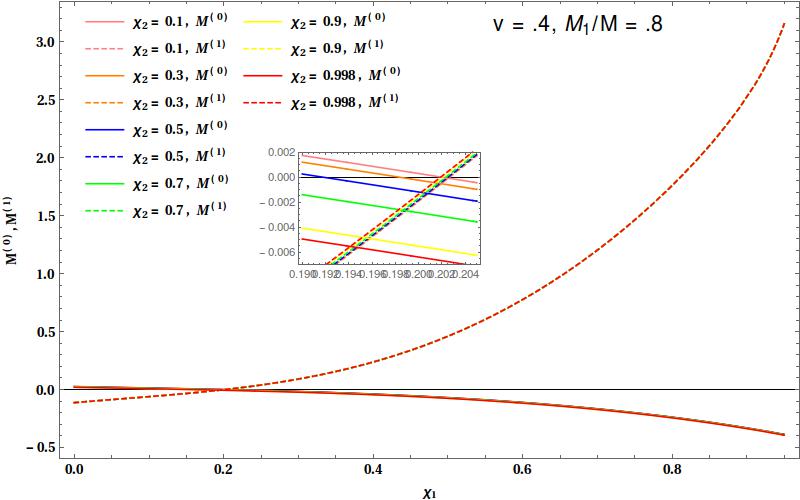}
\includegraphics[width=7.35cm]{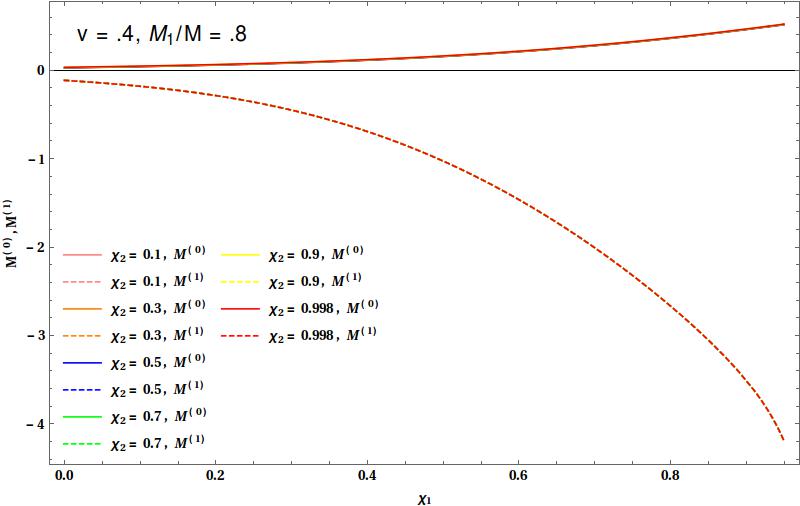}
\includegraphics[width=7.35cm]{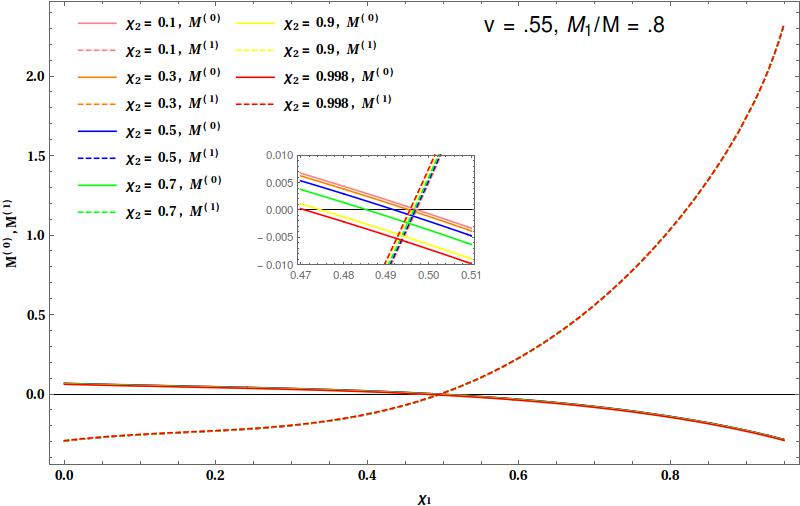}
\includegraphics[width=7.35cm]{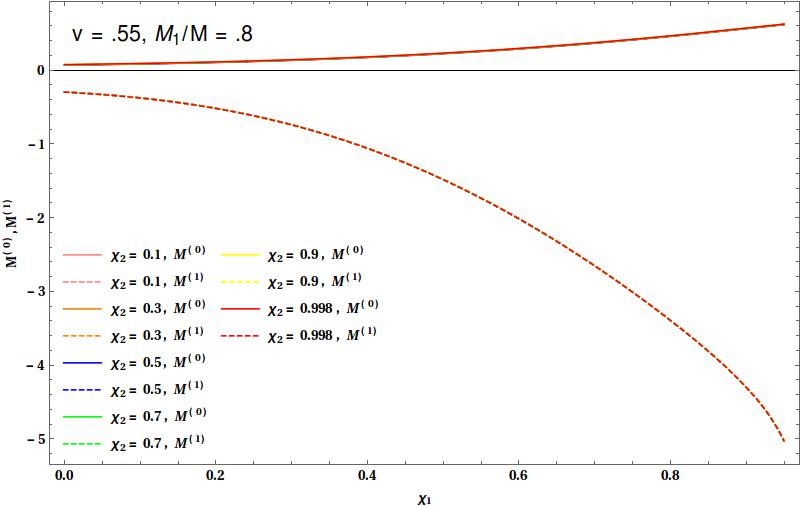}
\includegraphics[width=7.35cm]{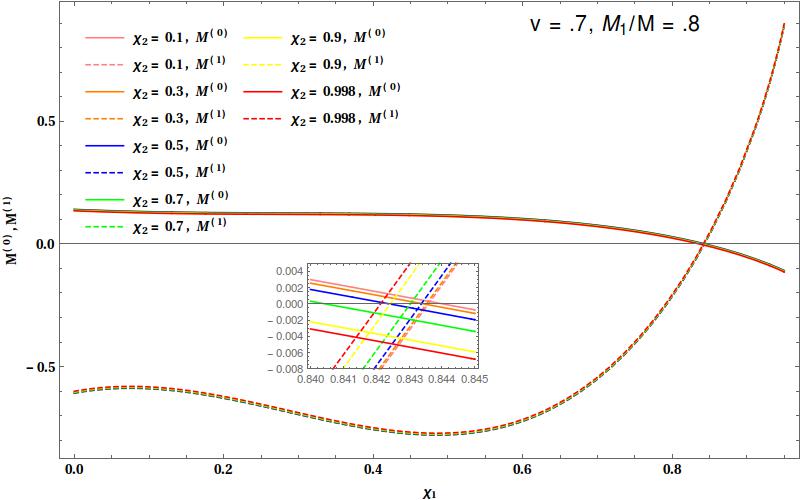}
\includegraphics[width=7.35cm]{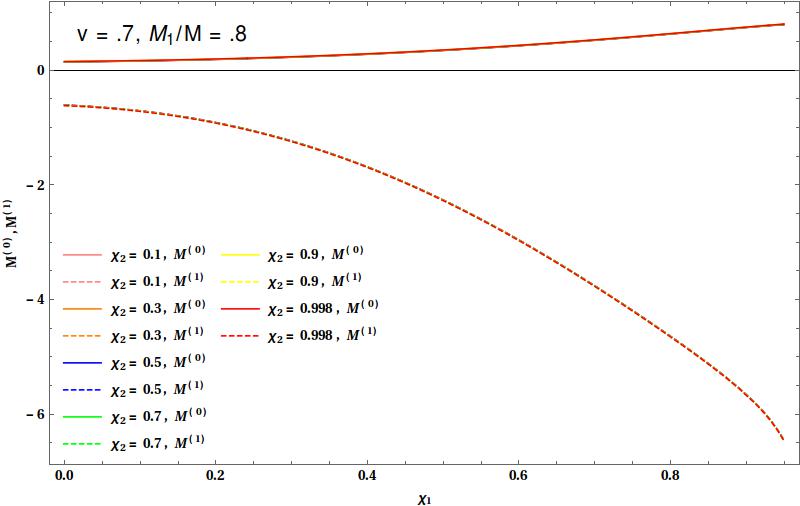}
\caption{${\rm M^{(0)}}$ and ${\rm M^{(1)}}$ has been plotted w.r.t the spins and the masses of the system. All the plots in the left panel represent systems in which the spin of the both of the components are aligned with the orbital angular momentum, whereas in the right panel they are anti-aligned. The mass ratio has been taken to be $M_1/M = .8$ for all the plots. Post-Newtonian velocity parameter $v$ has been taken to be $.4, .55$, and $.7$ for the plots in the first, second, and third row respectively.}
\label{q.8}
\end{figure*}
\end{widetext}

\begin{widetext}
\begin{figure*}
\includegraphics[width=7.35cm]{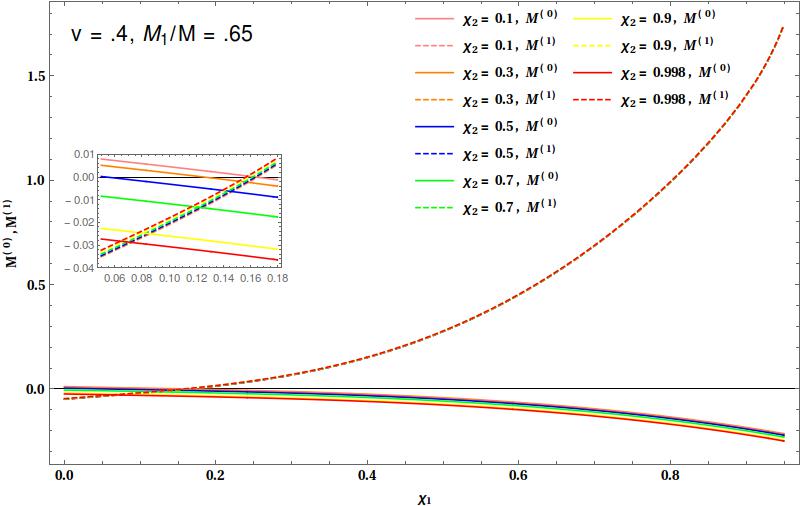}
\includegraphics[width=7.35cm]{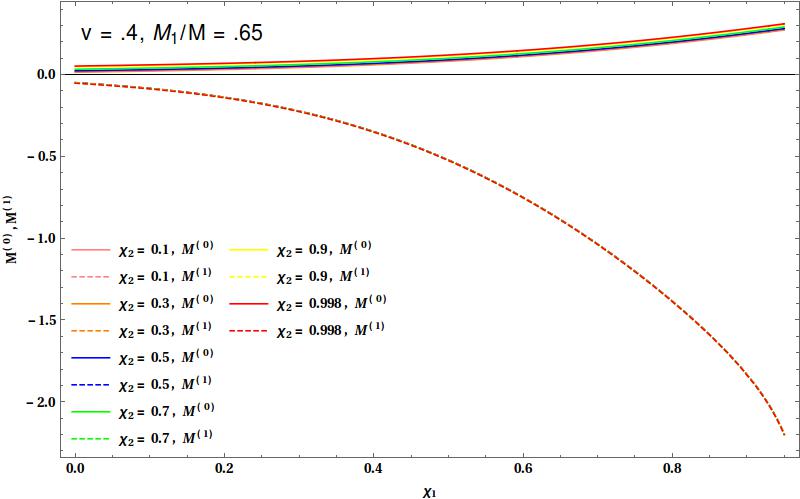}
\includegraphics[width=7.35cm]{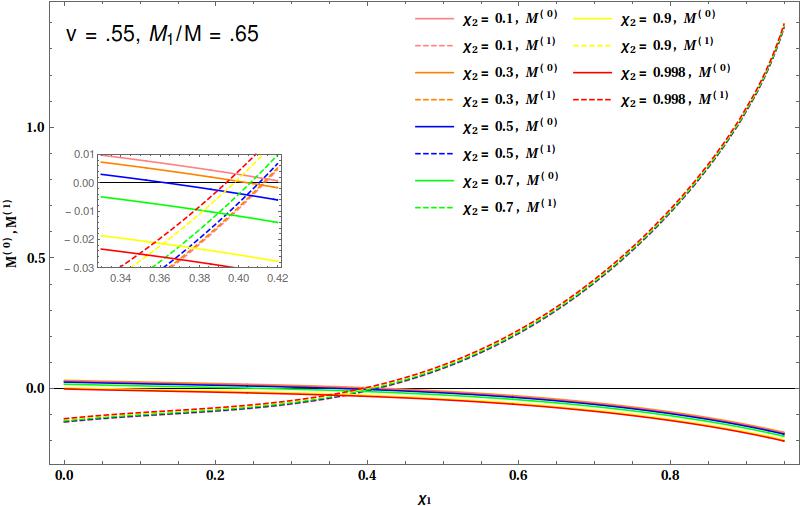}
\includegraphics[width=7.35cm]{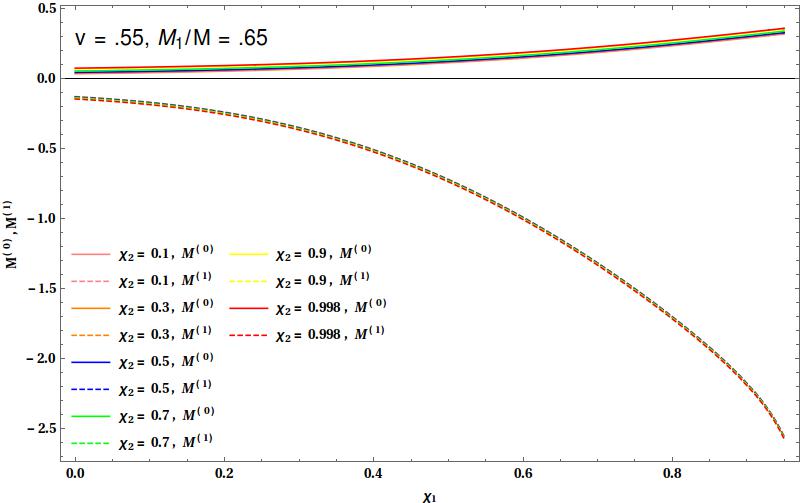}
\includegraphics[width=7.35cm]{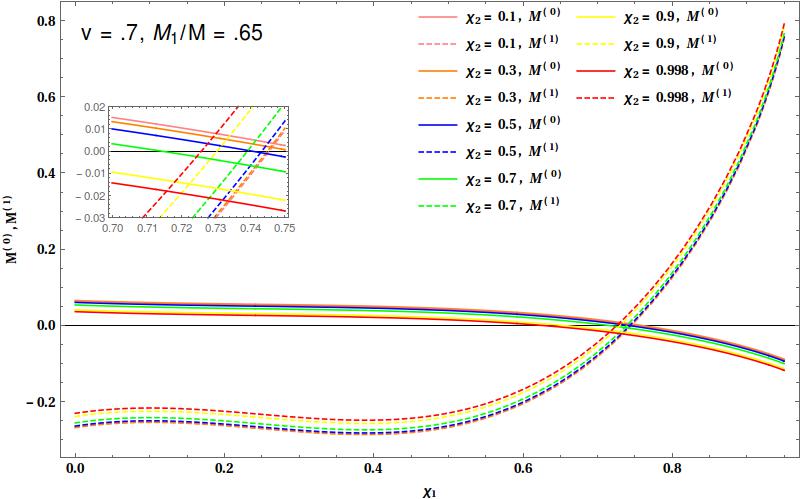}
\includegraphics[width=7.35cm]{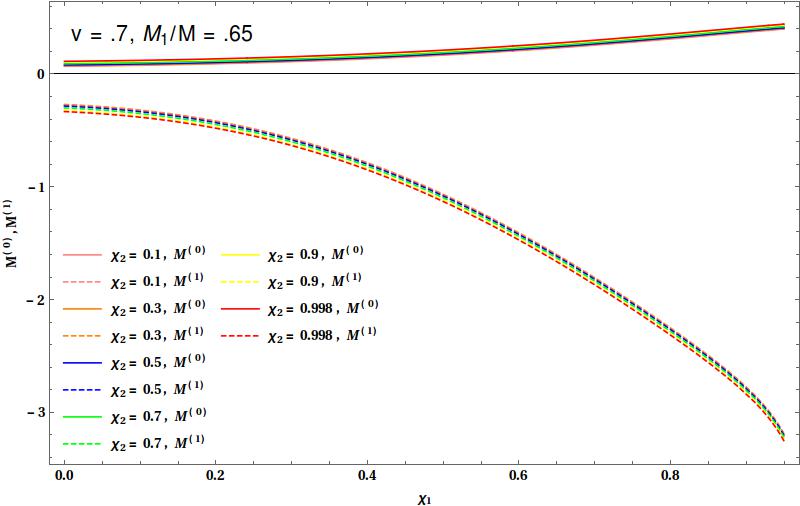}
\caption{${\rm M^{(0)}}$ and ${\rm M^{(1)}}$ has been plotted w.r.t the spins and the masses of the system. All the plots in the left panel represent systems in which the spin of the both of the components are aligned with the orbital angular momentum, whereas in the right panel they are anti-aligned. The mass ratio has been taken to be $M_1/M = .65$ for all the plots. Post-Newtonian velocity parameter $v$ has been taken to be $.4, .55$, and $.7$ for the plots in the first, second, and third row respectively.}
\label{q.65}
\end{figure*}
\end{widetext}

\begin{widetext}
\begin{figure*}
   \includegraphics[width=7.35cm]{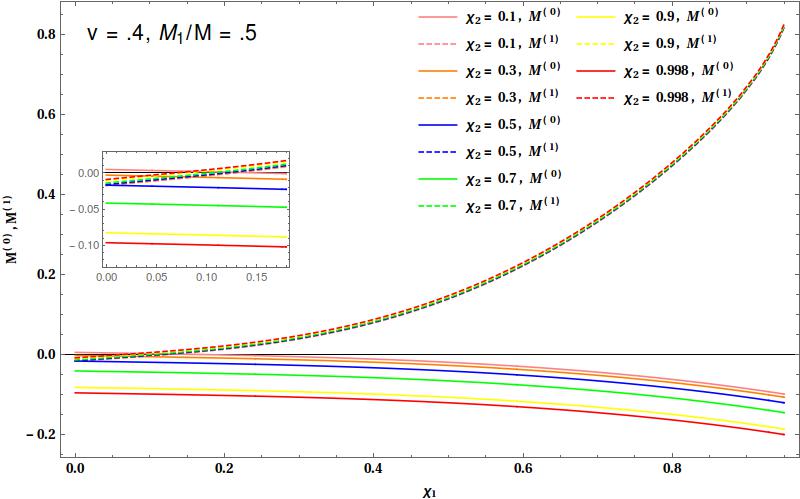}
\includegraphics[width=7.35cm]{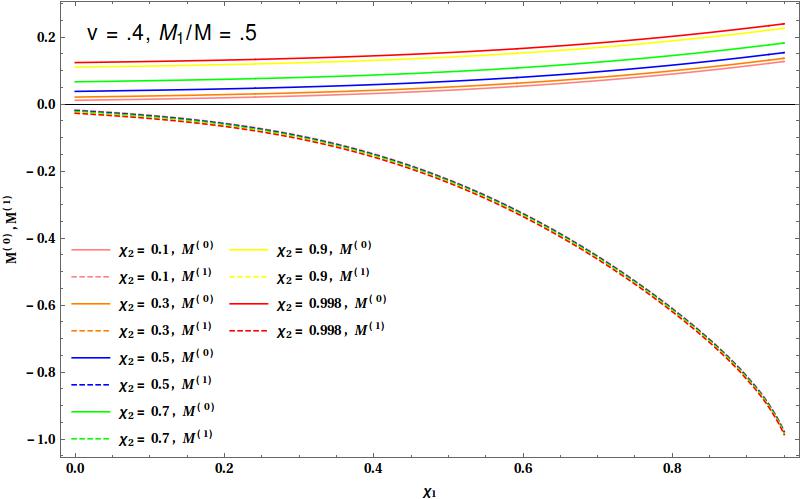}
\includegraphics[width=7.35cm]{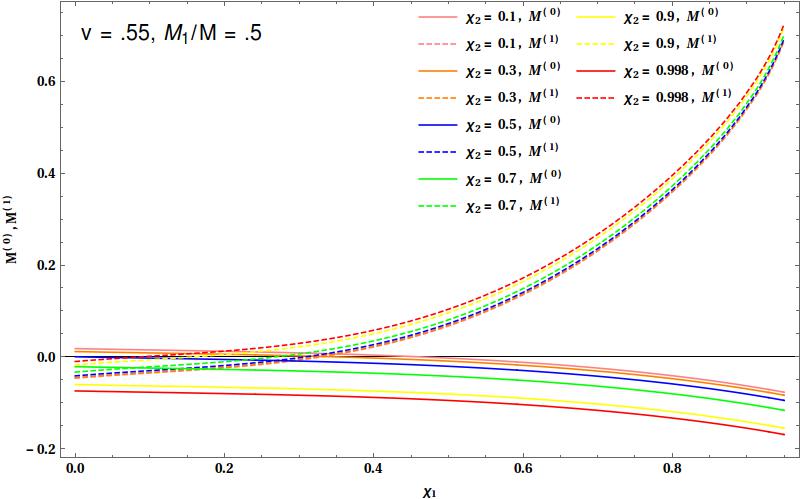}
\includegraphics[width=7.35cm]{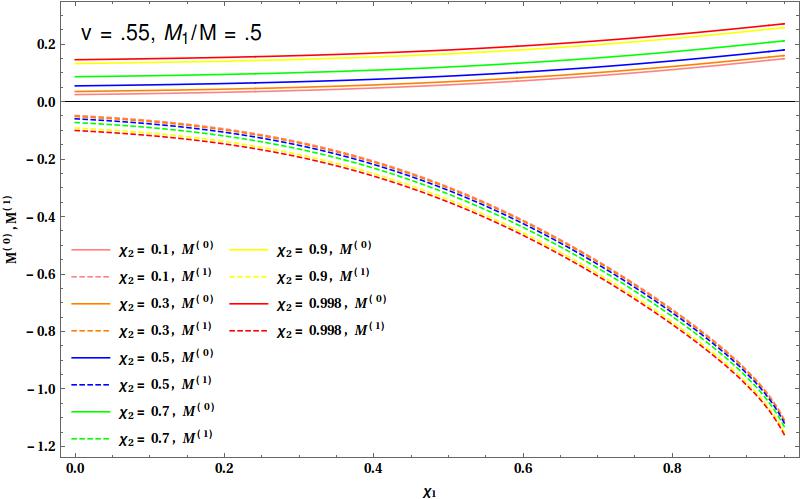}
\includegraphics[width=7.35cm]{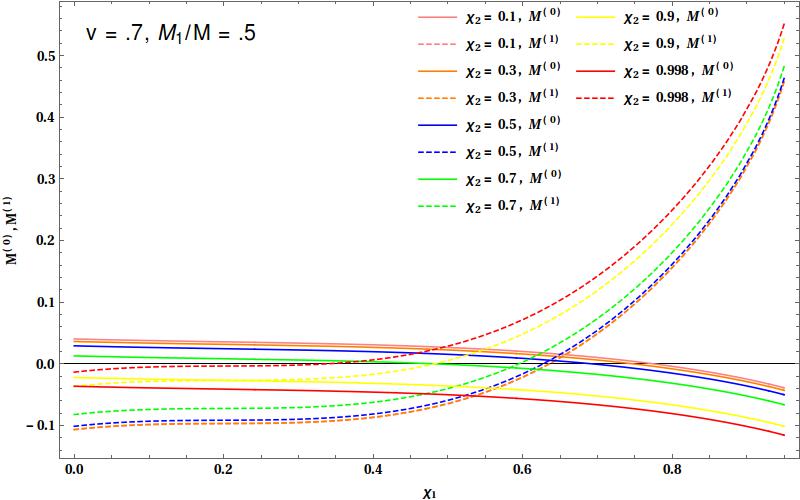}
\includegraphics[width=7.35cm]{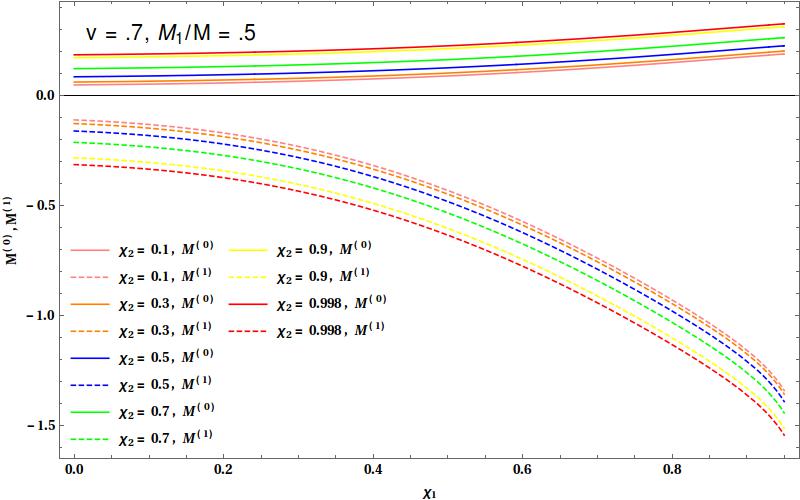}
\caption{${\rm M^{(0)}}$ and ${\rm M^{(1)}}$ has been plotted w.r.t the spins and the masses of the system. All the plots in the left panel represent systems in which the spin of the both of the components are aligned with the orbital angular momentum, whereas in the right panel they are anti-aligned. The mass ratio has been taken to be $M_1/M = .5$ for all the plots. Post-Newtonian velocity parameter $v$ has been taken to be $.4, .55$, and $.7$ for the plots in the first, second, and third row respectively.}
\label{q.5}
\end{figure*}
\end{widetext}

In Fig. \ref{q.8}, Fig. \ref{q.65} and Fig. \ref{q.5}, $M_1/M$ has been taken to be $.8, .65$ and $.5$ respectively. All of the figures show how ${\rm M^{(0)}}$ and ${\rm M^{(1)}}$ depend on the various parameters. All the plots in the left panel represent systems in which the spin of the both of the components are aligned with the orbital angular momentum, whereas in the right panel they are anti-aligned. Post-Newtonian velocity parameter $v$ has been taken to be $.4, .55$, and $.7$ for the plots in the first, second, and third row respectively.

From the figures shown in this work it can be concluded that mostly $|{\rm M}^{(0)}| \leq |{\rm M}^{(1)}|$. This has very important significance for EMRI that will be observed with space-based Laser~Interferometer~Space~Antenna~(LISA)~\cite{Audley:2017drz}. Since $|{\rm M}^{(0)}| \sim |{\rm M}^{(1)}|$, the value of $|\mathcal{R}|^2/\varepsilon$ will determine which one is more significant as far as the observation is concerned.

Approaching in an agnostic manner, this means that measuring $|\mathcal{R}|^2$ will be much more troublesome than anticipated. As both  $|\mathcal{R}|^2$ and $\varepsilon$ contribute in the same post-Newtonian order, there will be a degeneracy between $|\mathcal{R}|^2$ and $\varepsilon$ during parameter estimation. But the factor in front of $|\mathcal{R}|^2$ and $\varepsilon$, respectively $\mathcal{M}^{(0)}$ and $\mathcal{M}^{(1)}$, has different spin dependence. This may break the degeneracy partially but it is unlikely that it will be completely removed. Therefore from observation, we can find a joint posterior distribution of $|\mathcal{R}|^2$ and $\varepsilon$. We can also marginalize one parameter to find some estimates of the other one. But due to the degeneracy, it remains to see how well that estimation will be.

As long as EMRI is concerned, even though a detailed numerical analysis is needed to comment on the observability of $\varepsilon$, it is possible to do an order of magnitude estimation. We will do it by using available results in the literature. In \cite{Datta:2019epe} it has been shown that $|\mathcal{R}|^2$ can be constrained down to the value $\leq 5 \times 10^{-5}$ for SNR $(\rho) \sim 20$ with a $\chi_1 = .8$ for the supermassive body in the EMRI.

Two waveforms are considered indistinguishable for parameter estimation purposes if mismatch $\mathfrak{M} \lesssim 1/(2\rho^2)$~\cite{Flanagan:1997kp,Lindblom:2008cm} (for definition check appendix \ref{mismatch}), where $\rho$ is the SNR of the true
signal. For an EMRI with an SNR $\rho \approx 20$ (resp., $\rho \approx 100)$ one has $\mathfrak{M}  \lesssim 10^{-3}$ (resp., $\mathfrak{M} \lesssim 5 \times 10^{-5})$. Ref. \cite{Datta:2019epe} showed that, considering a supermassive object with $\chi_1 \gtrsim 0.8$ and a signal with $\rho  = 20$, very stringent bound on the reflectivity $|\mathcal{R}|^2  \lesssim 5 \times 10^{-5}$ can be out with LISA. Requiring that the dephasing to be smaller than 1 rad. and considering also $\chi_1 \gtrsim 0.8$, a slightly weaker constraint $|\mathcal{R}|^2 \lesssim 10^{-4}$ can be put.

This analysis in Ref. \cite{Datta:2019epe} was done with a detailed numerical simulation but the assumption was that the rate of change of mass due to tidal heating is $\propto (\mathcal{M}^{(0)} -|\mathcal{R}|^2 \mathcal{M}^{(0)})$. By varying the values of $|\mathcal{R}|^2$ and calculating mismatch with the $|\mathcal{R}|^2=0$ (classical BH) case the conclusions were found in that work. 
We want to use that result to comment on the impact of $\varepsilon$ on the waveform. 

The conclusion regarding the constraints on the $|\mathcal{R}|^2$ was reached using the terms $|\mathcal{R}|^2 {\rm M}^{(0)}$. Since ${\rm M}^{(0)} \sim {\rm M}^{(1)}$, similar kind of conclusion can be reached for $\varepsilon {\rm M}^{(1)}$.
 If we assume that tidal heating is $\propto (\mathcal{M}^{(0)} -\varepsilon \mathcal{M}^{(1)})$ and replace $\mathcal{M}^{(1)}$ with $\mathcal{M}^{(0)}$ (since ${\rm M}^{(0)} \sim {\rm M}^{(1)}$), then the conclusions regarding $|\mathcal{R}|^2$ can be translated to $\varepsilon$. The estimation from this will be a conservative estimation, since $|{\rm M}^{(0)}| < |{\rm M}^{(1)}|$. Conclusion of this is presented in the next paragraph.

Considering a supermassive object with $\chi_1 \sim 0.8$ and a signal with $\rho  = 20$, this implies that even values as small as $\varepsilon  \lesssim 5 \times 10^{-5}$ can have observable $\mathfrak{M}\gtrsim 10^{-3}$. For increased SNR $(\rho \sim 100)$ even smaller values of $ \varepsilon$ can have observational impact (such as $\mathfrak{M} \sim 10^{-5}$)\footnote{This comment is not entirely accurate since spin dependence of ${\rm M}^{(0)}$ and ${\rm M}^{(1)}$ are different. Nevertheless as an order of magnitude estimation this result is important.}. It is likely that much more stringent constraints can be found in reality, since $|{\rm M}^{(0)}| < |{\rm M}^{(1)}|$.

\begin{widetext}
\begin{figure*}
\includegraphics[width=7.35cm]{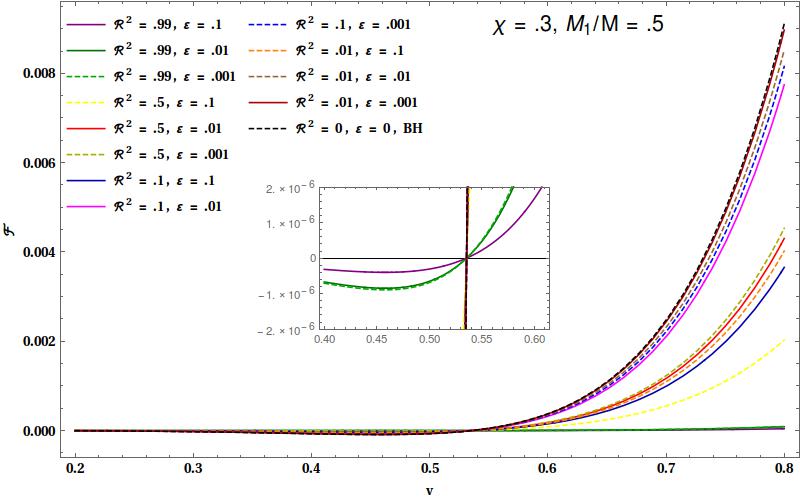}
\includegraphics[width=7.35cm]{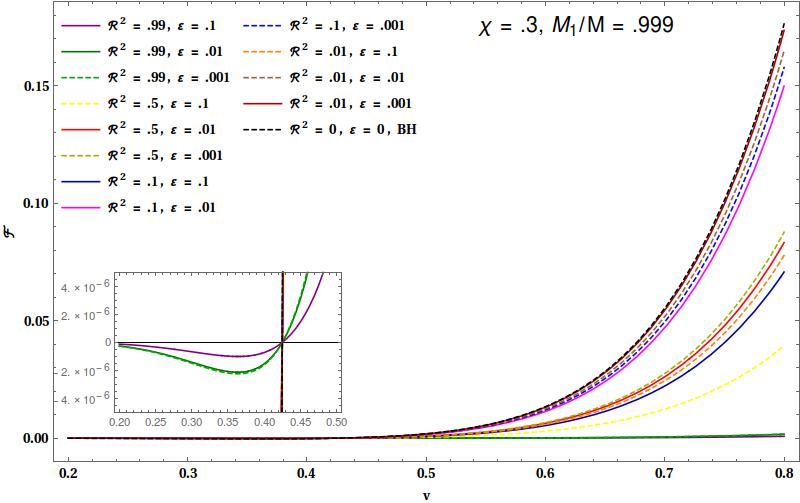}
\includegraphics[width=7.35cm]{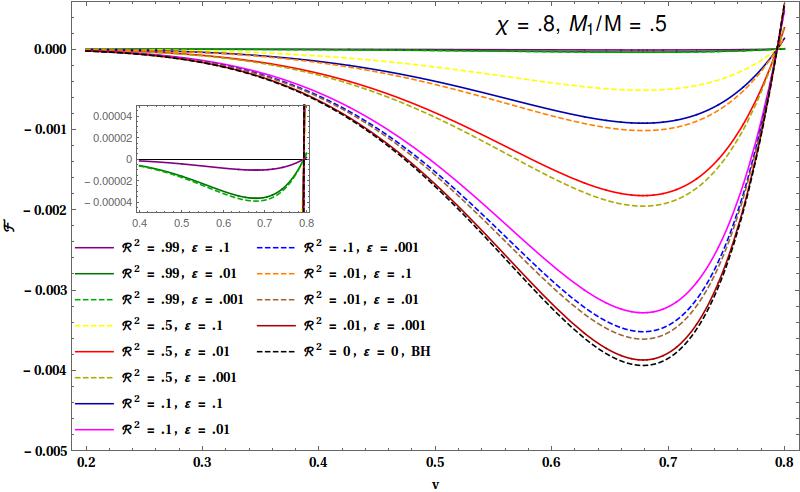}
\includegraphics[width=7.35cm]{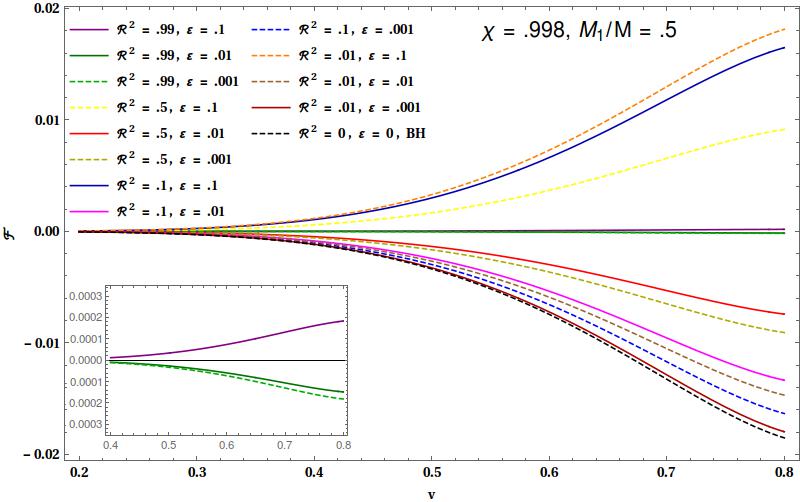}
\includegraphics[width=7.35cm]{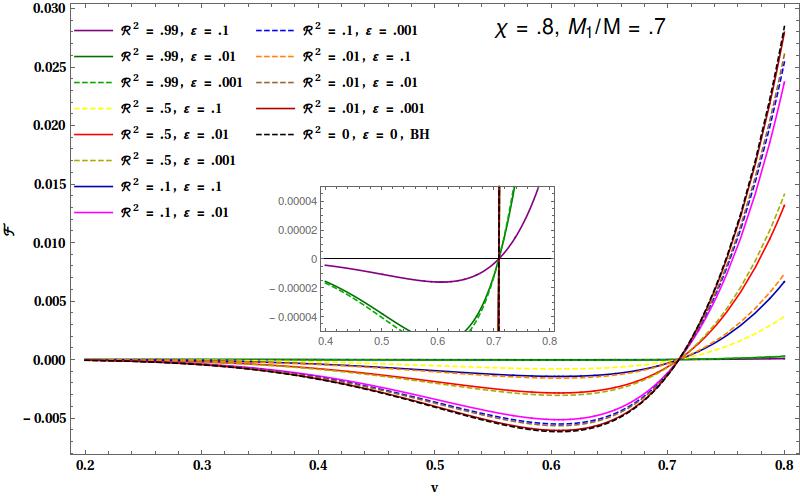}
\includegraphics[width=7.35cm]{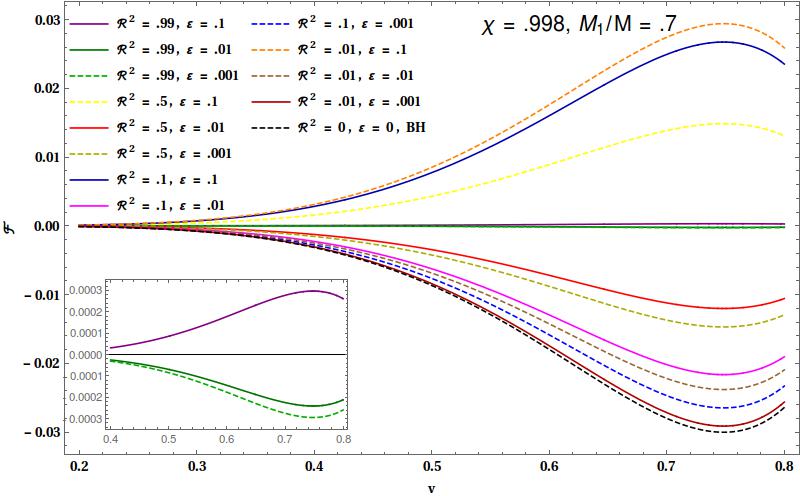}
\includegraphics[width=7.35cm]{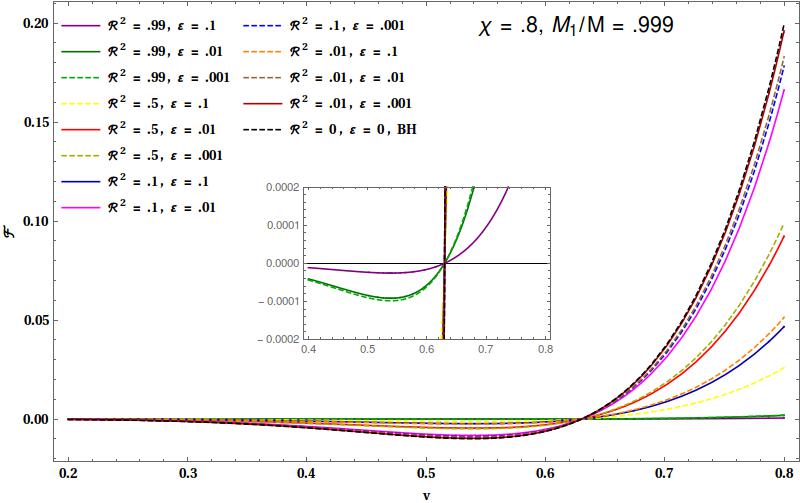}
\includegraphics[width=7.35cm]{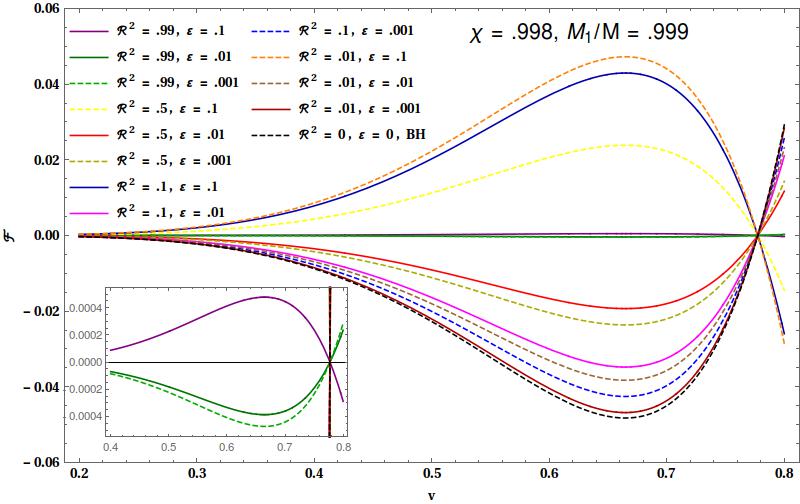}
\caption{$\mathcal{F}$ has been plotted w.r.t. PN velocity parameter $v$. For all the plots $|\mathcal{R}|^2$ and $\varepsilon$ is kept fixed and the spin and the mass ratio have been varied. All the plots were constructed by taking the spin of the KECO to be parallel to the orbital angular momentum.}
\label{F1}
\end{figure*}
\end{widetext}

\subsection{Superradiance}

Superradiance is the phenomenon when the energy is lost from the body \cite{Zel'dovich_TH, Misner:1972kx, Press:1972zz, Vicente:2018mxl, Brito:2015oca}. In case of a BH (KECO) this implies when $\frac{dM}{dt}<0$. From the results earlier we can write $\frac{dM}{dt} \propto (\mathcal{M}^{(0)} -|\mathcal{R}|^2 \mathcal{M}^{(0)} + \varepsilon \mathcal{M}^{(1)})$. It is possible to have $\frac{dM}{dt} < 0$ even though $\mathcal{M}^{(0)} >0$. This implies that the superradiance behavior of a KECO can be different (depending on the values of $|\mathcal{R}|^2$ and $\varepsilon$) from a BH of similar mass and spin. 

\begin{widetext}
\begin{figure*}
\includegraphics[width=7.35cm]{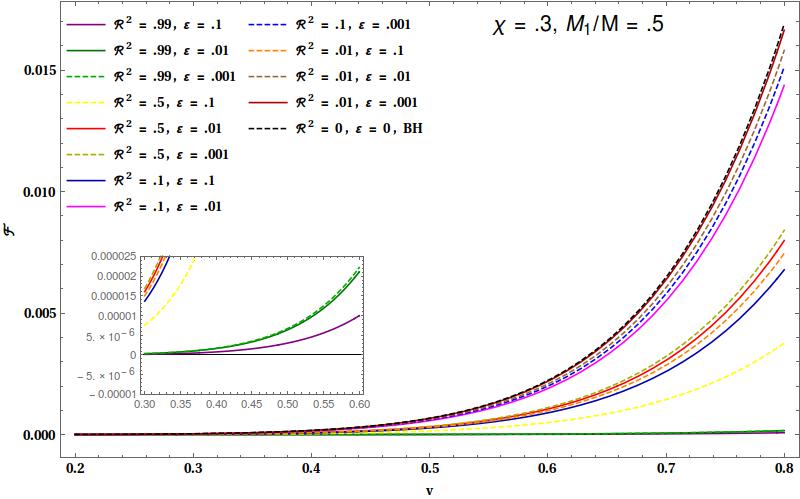}
\includegraphics[width=7.35cm]{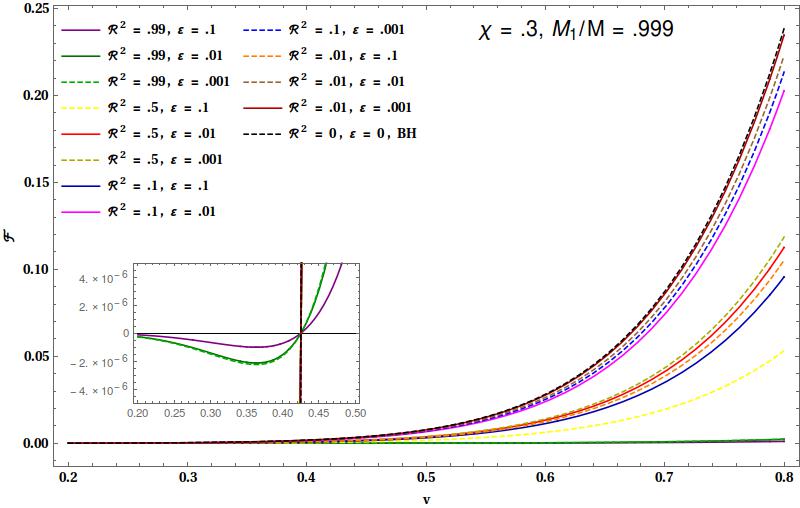}
\includegraphics[width=7.35cm]{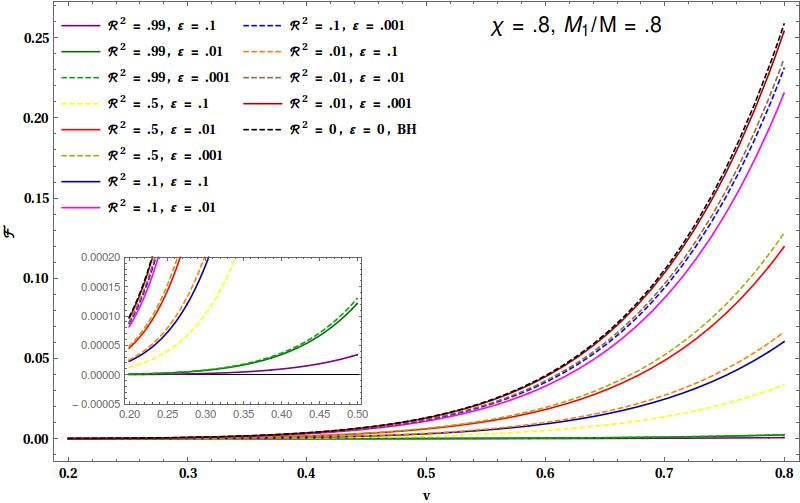}
\includegraphics[width=7.35cm]{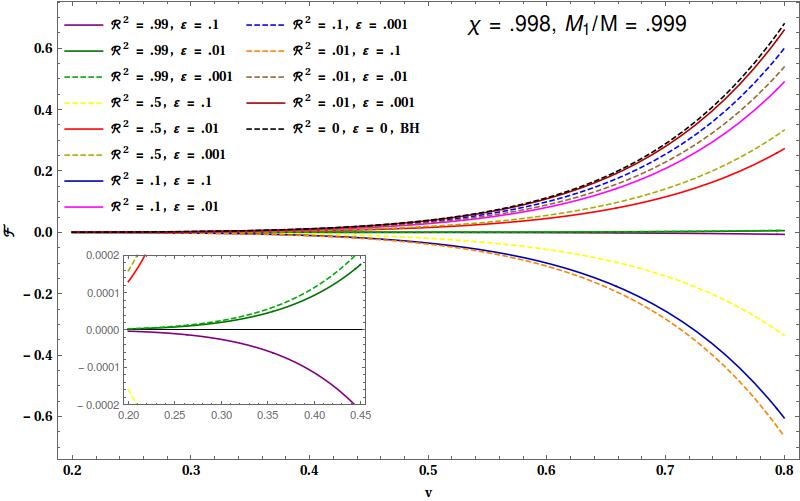}
\caption{$\mathcal{F}$ has been plotted w.r.t. PN velocity parameter $v$. For all the plots $|\mathcal{R}|^2$ and $\varepsilon$ is kept fixed and the spin and the mass ratio have been varied. All the plots were constructed by taking the spin of the KECO to be anti-parallel to the orbital angular momentum.}
\label{F-1}
\end{figure*}
\end{widetext}

To understand the phenomenon we compare the $\frac{dM_1}{dt}$ for a varied range of parameters. We define $\mathcal{F}$ as the fractional flux due to TH as follows:

\begin{equation}
    \mathcal{F} = \frac{dM_1}{dt}/\Big(\frac{dE}{dt}\Big)_N,
\end{equation}

where $\Big(\frac{dE}{dt}\Big)_N$ is the leading order flux at infinity defined in Eq.(\ref{flux at infinity}). In Fig. \ref{F1} and Fig. \ref{F-1} we plot $\mathcal{F}$ w.r.t. the PN velocity parameter $v$. The spin of the body has been taken to be aligned with the orbital angular momentum in In Fig. \ref{F1} whereas it is anti-aligned in In Fig. \ref{F-1}. The black dashed curve in both cases represents BH $(|\mathcal{R}|^2 =0, \varepsilon = 0)$. Other curves represent $(|\mathcal{R}|^2 \neq 0, \varepsilon \neq 0)$. Depending on the values of $|\mathcal{R}|^2$ and $\varepsilon$, the sign of $\mathcal{F}$ can be opposite of the BH. This will result in the presence (absence) of super-radiance in a parameter range where it is absent (present) for a BH.

\section{discussion}
\label{discussion}

We studied the tidal heating of an ECO that has a reflective surface at $r=r_+(1+\varepsilon)$. The metric outside the reflective surface has been considered to be that of the Kerr metric. We studied the tidal heating of such an object in the presence of a stationary companion. We showed that in stationary case energy dissipation through the reflective surface is zero similar to a Kerr BH. We calculated the rate of change of area and spin of such ECOs and showed that it depends on the position of the reflective surface. 

We also computed the tidal heating when such ECOs are in an inspiralling binary. Here the rate of change of mass, spin, and area of the ECO is different from a BH and depends on the position of the reflective surface. In the BH limit $(\varepsilon \rightarrow 0, {\cal R} \rightarrow 0, {\mathfrak T} \rightarrow 1)$ BH results are recovered. As a result, the phase of the GW emitted from the inspiring ECOs differs from inspiraling BBH not only because of the nonzero reflectivity but also due to nonzero $\varepsilon$. We found that all relevant quantities depend on $\varepsilon$ perturbatively, resulting in a series expansion in the powers of $\varepsilon$. We also discussed the potential degeneracy between $|\mathcal{R}|^2$ and $\varepsilon$, in the context of observation.

Point to note is, we achieved this with minimal assumptions. In our approach we were conservative. ECOs considered in the current work differs from Kerr BH only due to the presence of the reflective surface. Details of the interior of the ECO is not very important for our purpose. Metric outside the surface matches that of a Kerr metric. The main approach that we have followed here will be valid for almost every kind of ECOs. The main changes that will arise are discussed below:

\begin{itemize}
\item \emph{Surface geometry} of different kinds of ECOs can modify Eq.~(\ref{area rate of QBH}).
\item \emph{Non-Kerr metric outside of the surface} will modify the perturbation equations of the metric. This will change the functions in Eq.~(\ref{R_m}).
\item \emph{Nonzero energy-momentum tensor} will modify Eq.(\ref{NP sigma eqn main text}) (i.e. matter fields) along with the next-to leading order modification described in the last paragraph of appendix \ref{NP appendix}. 
\end{itemize}

In this work, we focused on the terms that are $\mathcal{O}(\varepsilon)$ and that has $|\mathcal{R}|^2$ reflectivity dependence. But it is easy to see that there will be terms like $\mathfrak{T}^*\mathcal{R}$ and $\mathfrak{T}\mathcal{R}^*$. Assuming $\mathfrak{T}$ and $\mathcal{R}$ to be real quantities and $\mathfrak{T}^2 = 1-|\mathcal{R}|^2$, this implies that there will be contribution $\sim \mathcal{R}\sqrt{1-|\mathcal{R}|^2}$. 

Another important point that needs to be addressed is the frequency dependence of $\mathcal{R}$. In this work, we did not address what kind of results we should expect in the presence of frequency dependence. For a short discussion check appendix \ref{Frequency dependent reflectivity}.

These points would be the center of the investigation in the current future. The results found in this work shows specifically that the modification of the horizon geometry not only brings reflectivity but also $\varepsilon$ in the observable footing, even in the inspiral phase of a binary. This brings us the possibility to test the nature of the surface of binary components using GW.

\section*{acknowledgement}

I would like to thank Sukanta Bose for his continuous guidance. I would like to thank Richard Brito, Vitor Cardoso, Elisa Maggio, Paolo Pani, and Karthik Rajeev for useful discussions. I thank Dipanjan Rai Chaudhuri for his support and for increasing my curiosity. I would like to thank the University Grants Commission (UGC), India, for financial support as a senior research fellow. 

\appendix

\section{Coefficients of the expansions}
\label{Coefficients}


\begin{widetext}
\begin{equation}
\dot{A}^{(0)} =-\frac{8 \pi  M_1^5 M_2^2 s^2 \sin ^2(\theta _0)  \Big(15 s^2 \cos (2 \theta _0)-9 s^2-8\Big)}{5   b^6 \sqrt{1-s^2}}
\end{equation}

\begin{equation}
\begin{split}
\dot{A}^{(1)} =&\frac{8 \pi   M_1^5 M_2^2 \chi_1^4  \sin ^2(\theta _0)}{35 b^6 \Big(\chi_1^2-1\Big) \Big(\sqrt{1-\chi_1^2}+1\Big)^2} \Big[\Big\{\big(-84\chi_1^6 + 294\chi_1^4 + 359 \chi_1^2 - 674\big) + \sqrt{1-\chi_1^2}\big(231\chi_1^4 + 23 \chi_1^2 -674\big)\Big\} \cos (2 \theta _0)\\
&+(-84\chi_1^6 + 650 \chi_1^4 - 1913 \chi_1^2 +1466) + \sqrt{1-\chi_1^2}(195 \chi_1^4 -1185\chi_1^2 +1466)\Big]
\end{split}
\end{equation}


\begin{equation}
    \dot{S}^{(0)}_{\theta_0} = \frac{ M_1^5 M_2^2 \chi_1 \sin ^2(\theta _0)  \bigg(15 \chi_1^2 \cos (2 \theta _0)-9 \chi_1^2-8\bigg)}{5 b^6 }
\end{equation}

\begin{equation}
    \begin{split}
        \dot{S}^{(1)}_{\theta_0} =&\frac{M_2^2 M_1^5 \chi_1   \sin ^2(\theta _0)}{210 b^6
   \big(\sqrt{1-\chi_1^2}+1\big)^3  \big(1-\chi_1^2\big) } \Big[3\chi_1^2 \cos (2 \theta _0) \Big\{\big(168\chi_1^8 - 483\chi_1^6 - 4439 \chi_1^4 + 11090 \chi_1^2 - 6336) + \sqrt{1-\chi_1^2}(-525\chi_1^6\\
   &-1271\chi_1^4   + 7922\chi_1^2 -6336\big)\Big\} + \Big(504 \chi_1^{10}-6897\chi_1^8 + 30987 \chi_1^6 - 42354 \chi_1^4 + 11936\chi_1^2 + 5824\Big)+ \sqrt{1-\chi_1^2}\Big(-1863\chi_1^8\\ &+16107\chi_1^6-34202 \chi_1^4 + 14848 \chi_1^2 + 5824\Big) \Big]
    \end{split}
\end{equation}

\begin{eqnarray}
\dot{S}^{(0)}_{\frac{\pi}{2}} = &-\frac{8  M_1^5 M_2^2 \chi_1 \big(3 \chi_1^2+1\big) }{5 b^6 }\\
\dot{S}^{(1)}_{\frac{\pi}{2}} = &-\frac{4 M_1^5 M_2^2 \chi_1}{105 b^6    \big(1-\chi_1^2\big)  \big(1+
   \sqrt{1-\chi_1^2}\big)^3} \Big[(681\chi_1^8 - 5538\chi_1^6 + 7246 \chi_1^4 -3868\chi_1^2 -728) + \sqrt{1-\chi_1^2}(36\chi_1^8 -2490 \chi_1^6 +7246\chi_1^4\\\nonumber
   &-4232 \chi_1^2 -728)\Big]
\end{eqnarray}

\begin{equation}
    \begin{split}
        {\cal I}_0^{(0)} = &-\frac{2 M_1^6 M_2^2  \sin ^2(\theta _0)  \bigg(15 \chi_1^2 \cos (2 \theta _0)-9 \chi_1^2-8\bigg) [1+\sqrt{1-\chi_1^2}]}{5 b^6 }\\
{\cal I}_0^{(1)} = &-\frac{ M_2^2 M_1^6   \sin ^2(\theta _0)}{105 b^6 \big(\sqrt{1-\chi_1^2}+1\big)^2   \big(1-\chi_1^2\big) }  \bigg[ \bigg\{(168\chi_1^8 - 483\chi_1^6 -4439 \chi_1^4 + 11090 \chi_1^2 -6336) + \sqrt{1-\chi_1^2}(-525\chi_1^6 -1271 \chi_1^4 \\
&+ 7922 \chi_1^2 -6336)\bigg\}3 \chi_1^2 \cos (2
   \theta _0)+ \bigg\{(504 \chi_1^{10} - 6897 \chi_1^8 + 30987 \chi_1^6 - 42354 \chi_1^4+11936 \chi_1^2 + 5824) + \sqrt{1-\chi_1^2}(-1863\chi_1^8 \\
   &+ 16107 \chi_1^6-34202\chi_1^4 +14848\chi_1^2 + 5824)\bigg\}\bigg]
    \end{split}
\end{equation}

\begin{eqnarray}
{\cal S}^{(0)} = &\frac{v^5M_1^3}{4M^3 }(- \chi_1 +2\hat{L}_N.\hat{s}_1v^3\frac{M_1}{M}(1+\sqrt{1-\chi_1^2}))    \bigg[3 \chi_1^2+1\bigg]\\
{\cal S}^{(1)} = &\frac{ v^5 M_1^3}{4 M^3}   (- \chi_1 +2\hat{L}_N.\hat{s}_1v^3\frac{M_1}{M}(1+\sqrt{1-\chi_1^2})) \frac{\bigg[(681\chi_1^8 - 5538\chi_1^6 + 7246 \chi_1^4 -3868\chi_1^2 -728) + \sqrt{1-\chi_1^2}(36\chi_1^8 -2490 \chi_1^6 +7246\chi_1^4-4232 \chi_1^2 -728)\bigg]}{42\big(\sqrt{1-\chi_1^2}+1\big)^3  \big(1-\chi_1^2\big)}
\end{eqnarray}


\begin{eqnarray}
{\cal M}^{(0)} =& \hat{L}_N.\hat{s}_1{\cal S}^{(0)}.\\
{\cal M}^{(1)} = &\hat{L}_N.\hat{s}_1 {\cal S}^{(1)}.
\end{eqnarray}

\begin{eqnarray}
\label{psi0}
\psi^{(0)} = &\frac{40}{9} \bigg(8 \pi  \mathcal{M}_5^{(0)}-\mathcal{M}_8^{(0)}\bigg) v^8 (3 \log (v)-1)+\frac{5}{42} (952 \nu +995)\mathcal{M}_5^{(0)} v^7+\frac{40}{9} \mathcal{M}_5^{(0)} v^5 (3 \log (v)+1) +1 \leftrightarrow 2.\\
\label{psi1}
\psi^{(1)} = &\frac{40}{9} \bigg(8 \pi  \mathcal{M}_5^{(1)}-\mathcal{M}_8^{(1)}\bigg) v^8 (3 \log (v)-1)+\frac{5}{42} (952 \nu +995) \mathcal{M}_5^{(1)} v^7+\frac{40}{9} \mathcal{M}_5^{(1)} v^5 (3 \log (v)+1) + 1\leftrightarrow 2.
\end{eqnarray}

\end{widetext}

\section{Frequency dependent reflectivity}
\label{Frequency dependent reflectivity}

In this section, we will discuss the expected changes if the ${\mathfrak T}$ and $\mathcal{R}$ are frequency dependent. How these quantities will depend on the frequency depends specifically on the model under consideration. But it is always possible to write,
\begin{equation}
{\mathfrak T} (f) = {\mathfrak T}_0 {\cal{T}}(\frac{f}{f_0}),
\end{equation}

\begin{equation}
\mathcal{R} (f) = \mathcal{R}_0 R(\frac{f}{f_0}),
\end{equation}

where ${\cal{T}}$ and $R$ are some frequency-dependent functions but ${\mathfrak T}_0$ and $\mathcal{R}_0$
are frequency independent and $f_0$ has the dimension of frequency. For small frequency $(f \ll f_0)$, it is always possible to expand these functions as follows,
\begin{equation}
\mathcal{T}(\frac{f}{f_0}) = 1+\mathcal{T}'(0)\frac{f}{f_0} +\mathcal{T}''(0)\frac{f^2}{2f_0^2} +...
\end{equation}

\begin{equation}
R(\frac{f}{f_0}) = 1+R'(0)\frac{f}{f_0} +R''(0)\frac{f^2}{2f_0^2} +...
\end{equation}
where prime denotes the derivative w.r.t. the argument and $0$ inside the braces represent $f = 0$. For an inspiraling binary, we can identify this frequency with the frequency of the GW that is twice the frequency of the orbital motion $(\Omega)$. Therefore, we have $v^3 \propto \Omega \propto f$ where $v$ is the post Newtonian velocity parameter. So we can rewrite, 

\begin{equation}
\mathcal{T}(\frac{v}{v_0}) = 1+\mathcal{T}'(0)\frac{v^3}{v_0^3} +\mathcal{T}''(0)\frac{v^6}{2 v_0^6} +...
\end{equation}

\begin{equation}
R(\frac{v}{v_0}) = 1+R'(0)\frac{v^3}{v_0^3} +R''(0)\frac{v^6}{2 v_0^6} +...
\end{equation}

Hence upto $\mathcal{O}(v^3)$,

\begin{equation}
|{\mathfrak T} (v)|^2 = |{\mathfrak T}_0|^2 |1+2\mathcal{T}'(0)\frac{v^3}{v_0^3}| ,
\end{equation}

\begin{equation}
|\mathcal{R} (v)|^2 = |\mathcal{R}_0|^2 |1+2R'(0)\frac{v^3}{v_0^3}|.
\end{equation}

We have shown that the leading order reflectivity dependence arises at $2.5$ PN correction. Therefore the leading order contribution due to the frequency dependence will arise at 4 pn. For a model of a quantum black hole as discussed in Ref. \cite{Oshita:2019sat} this implies,

\begin{equation}
|\mathcal{R} (v)|^2 = |1-4\frac{\hbar}{k T_H}\frac{v^3}{G M c^3}|.
\end{equation}

\section{Discussion on integral $I$}
\label{horizon integral}

The horizon integral for BH has been discussed extensively in \cite{Thorne:1986iy}. The results are found explicitly for a source of tidal field orbiting the hole rigidly. Denote by $\Omega_m$ the common angular velocity (relative to distant inertial frames) of the source. In such case, $\phi$ and $t$ dependences of tidal fields will be,

\begin{equation}
{\rm all\,\,first\,\, order\,\, perturbation} = f(\phi -\Omega t) = f[\Bar{\phi} - (\Omega - \Omega_H)t].
\end{equation}

Due to this time derivative and $\phi$ derivative becomes connected via

\begin{equation}
\frac{\partial}{\partial t}\Big|_{\Bar{\theta}\Bar{\phi}} = -(\Omega-\Omega_H)\frac{\partial}{\partial\Bar{\phi}}\Big|_{t, \Bar{\theta}},
\end{equation}

$\Bar{\theta}$ and $\Bar{\phi}$ are the co-moving angular coordinates (for further details check Eq.(6.69), Eq.(7.19), and Eq.(7.21) of Ref. \cite{Thorne:1986iy}, where BH case has been explicitly derived). The point to be noted that the factor $(\Omega-\Omega_H)$ does not care about the properties of the "hole". This factor arises solely due to orbital motion. Therefore this should stay unchanged even if we replace the horizon with a reflective surface.

Finally it was shown
\begin{equation}
\frac{dJ}{dt} = (\Omega-\Omega_H) \oint_{\mathcal{H}} \frac{1}{8\pi}
\frac{\partial\Sigma^{\mathcal{H}}_{ab}}{\partial\Bar{\phi}}\frac{\partial\Sigma_{\mathcal{H}}^{ab}}{\partial\Bar{\phi}} dA.
\end{equation}

where $\mathcal{H}$ represents BH horizon and $\Sigma^{\mathcal{H}}_{ab}$ is related to the divergence $\sigma^{\mathcal{H}}_{ab}$ as follows

\begin{equation}
\sigma^{\mathcal{H}}_{ab} = \frac{\partial \Sigma^{\mathcal{H}}_{ab}}{\partial t}
\end{equation}

Therefore for KECO's reflective surface we will have,

\begin{equation}
\frac{dJ}{dt} = (\Omega-\Omega_H)I = (\Omega-\Omega_H)\Big[ \oint_{\mathcal{H}} \frac{1}{8\pi}
\frac{\partial\Sigma^{\mathcal{H}}_{ab}}{\partial\Bar{\phi}}\frac{\partial\Sigma_{\mathcal{H}}^{ab}}{\partial\Bar{\phi}} dA + \mathcal{O}(\varepsilon)\Big],
\end{equation}

where the first part is the BH result and there will $\mathcal{O}(\varepsilon)$ correction due to KECO.

$\Omega \propto v^3$, therefore the leading order PN correction will arise from $(\Omega-\Omega_H)$. Since $v$ and therefore $\Omega$ is a small quantity it is possible to expand $I$ in an expansion of $\Omega$. Therefore the leading order PN correction would be $=(\Omega-\Omega_H)I(\Omega = 0)$. In case of a stationary source this result with $(\Omega-\Omega_H)|_{\Omega = 0} = -\Omega_H$ will be valid. but the result for the stationary case has already been found explicitly in this paper. Therefore only thing remains is to identify $I(\Omega = 0)$ in that result, which can be done by comparing with Eq. (\ref{stationary area rate}) and Eq. (\ref{stationary s rate}) along the line of Ref. \cite{Alvi:2001mx}. The first part will give the BH result $\mathcal{O}(\varepsilon^0)$ and the $\mathcal{O}(\varepsilon)$ part will give the $\varepsilon$ dependent contribution. This second part can be expanded in the series expansion of $\varepsilon$. But the crucial point is, as long as the leading order pn terms are concerned, we can identify the results by taking the stationary limit rather than explicitly evaluating the integral.

\section{Mismatch}
\label{mismatch}

To assess whether an effect is sufficiently strong to be measurable in a GW detector with  noise power spectral density $S_n(f)$, is to compute the overlap $\mathcal{O}$ between two waveforms $h_1(t)$ and $h_2(t)$:
\begin{equation}\label{overlap}
\mathcal{O}(h_1|h_2) = \frac{\left\langle h_1|h_2\right\rangle}{\sqrt{\left\langle h_1|h_1\right\rangle \left\langle h_2|h_2\right\rangle}}\,,
\end{equation}
where, the inner product $\left\langle h_1|h_2\right\rangle$ is defined by
\begin{equation}
\left\langle h_1|h_2\right\rangle = 4\Re\,\int_{0}^{\infty} \frac{\tilde{h}_1 \tilde{h}^*_2}{S_n(f)} df\,.
\end{equation}
The tilded quantities stand for the Fourier transform and the star for complex conjugation. Since the waveforms 
are defined up to an arbitrary time and phase shift, it is necessary to maximize the overlap~\eqref{overlap} over 
these quantities. This can be done by computing~\cite{Allen:2005fk} 
\begin{equation}\label{overlap2}
\mathcal{O}(h_1|h_2) = \frac{4}{\sqrt{\left\langle h_1|h_1\right\rangle \left\langle h_2|h_2\right\rangle}}\max_{t_0} \left|\mathcal{F}^{-1}\left[\frac{\tilde{h}_1 \tilde{h}^*_2}{S_n(f)}\right](t_0)\right|\,,
\end{equation}
where $\mathcal{F}^{-1}[g(f)](t) =\int_{-\infty}^{+\infty} g(f) e^{-2\pi i f t}df$ represents the inverse Fourier 
transform. The overlap is defined such that $\mathcal{O}=1$ indicates a perfect agreement between the two waveforms. The mismatch $(\mathfrak{M})$ is defined as follows:

\begin{equation}
    \mathfrak{M}\equiv 1-{\mathcal{O}}
\end{equation}

\section{Newman-Penrose formalism}
\label{NP appendix}

\subsection{Basic definitions}

The geometry of space-time and its dynamics can be cast in a different form by defining a set of tetrads. In our  four-dimensional Riemannian space a tetrad system of vectors $l_{\mu}, m_{\mu}, \Bar{m}_{\mu}$, and $n_{\mu}$ can be introduced. $l_{\mu}$ and $n_{\mu}$ are real null vectors and $m_{\mu}$ and its complex conjugate$\Bar{m}_{\mu}$ are complex null vectors. The orthogonality properties of the vectors are,

\begin{equation}
    \begin{split}
        l_{\mu}l^{\mu} =& m_{\mu}m^{\mu} = \Bar{m}_{\mu}\Bar{m}^{\mu} = n_{\mu}n^{\mu} =0,\\
        l_{\mu}n^{\mu} =& -m_{\mu}\Bar{m}^{\mu} = 1,\\
        l_{\mu}m^{\mu} =& l_{\mu}\Bar{m}^{\mu} = n_{\mu}m^{\mu} = n_{\mu}\Bar{m}^{\mu} = 0.
    \end{split}
\end{equation}

Spin coefficients can be defined from this set of tetrads as follows:

\begin{eqnarray}
\kappa =& l_{\mu;\nu}m^{\mu}l^{\nu},\,\,\,\pi = -n_{\mu;\nu}\Bar{m}^{\mu}l^{\nu}\\
\epsilon =& \frac{1}{2}(l_{\mu;\nu}n^{\mu}l^{\nu} - m_{\mu;\nu}\Bar{m}^{\mu}l^{\nu}),\,\,\,\rho = l_{\mu;\nu}\Bar{m}^{\mu}\Bar{m}^{\nu}\\
\alpha =& \frac{1}{2}(l_{\mu;\nu}n^{\mu}\Bar{m}^{\nu} - m_{\mu;\nu}\Bar{m}^{\mu}\Bar{m}^{\nu}),\,\,\,\lambda = -n_{\mu;\nu}\Bar{m}^{\mu}\Bar{m}^{\nu}\\
\sigma =& l_{\mu;\nu}m^{\mu}m^{\nu},\,\,\,\mu = -n_{\mu;\nu}\Bar{m}^{\mu}m^{\nu}\\
\beta =& \frac{1}{2}(l_{\mu;\nu}n^{\mu}m^{\nu} - m_{\mu;\nu}\Bar{m}^{\mu}m^{\nu}),\,\,\,\nu = -n_{\mu;\nu}\Bar{m}^{\mu}n^{\nu}.
\end{eqnarray}

An overbar in this section implies complex conjugation. Equation satisfied by $\sigma$ is as follows:

\begin{equation}
        \label{eqn for spin coef}
    \begin{split}
D\sigma - \delta \kappa =& (\rho + \Bar{\rho})\sigma + (3\epsilon - \Bar{\epsilon})\sigma - (\tau - \Bar{\pi} + \Bar{\alpha} + 3\beta)\kappa\end{split}
\end{equation}

$D,\,\,\delta$ are derivative operators defined as follows,

\begin{equation}
   D = l^{\mu}\frac{\partial}{\partial x^{\mu}},\,\,\,\,\delta = m^{\mu}\frac{\partial}{\partial x^{\mu}}. 
\end{equation}

For further details check Ref. \cite{MTB}.

\subsection{Hartle-Hawking tetrad}
\label{Hartle-Hawking tetrad}
Hartle-Hawking (HH) tetrad is an useful tetrad for studying the properties of spacee-time near black holes \cite{Hawking:1972hy}. In Boyer-Lindquist co-ordinate systems the components are as follows:

\begin{equation}
\begin{split}
l^{\mu} =& [1/2, \Delta/2(r^2+a^2, 0, a/2(r^2+a^2))]\\
n^{\mu} =& [r^2+a^2, -\Delta, 0, a]\frac{(r^2 +a^2)}{\Delta \Sigma}\\
m^{\mu} =& [ia\sin\theta, 0, 1, i/\sin\theta]\frac{1}{\sqrt{2}(r+ia\cos\theta)}.
\end{split}
\end{equation}

In HH tetrad Eq.(\ref{eqn for spin coef}) simplifies to
\begin{equation}
\label{leading NP eqn}
D\sigma^{HH} = 2\epsilon\sigma^{HH} +\psi_0^{HH}
\end{equation}
for a black hole horizon \cite{Teukolsky:1974yv}. In the case of KECO, this equation will be satisfied only approximately. Since $\varepsilon << 1$, in the leading order Eq.(\ref{leading NP eqn}) will be valid. We will not investigate the modification of Eq.(\ref{leading NP eqn}) and its contribution to our final result. We will investigate the effect of these corrections and other assumptions made in the paper in another project in the current future. Even though without such modifications our final result is incomplete, nevertheless the methods described in this paper are very crucial, as it brings several disconnected pieces together, opening up a new research direction for the tidal heating of KECO.

\section{Teukolsky equation and its solutions}
\label{apndx:Teukolsky equation}
The equation satisfied by Weyl scalar $\psi_0$ in the vaccum is as follows \cite{Teukolsky:1973ha}:

\begin{widetext}
\begin{equation}
    \begin{split}
        \Big[\frac{(r^2+a^2)^2}{\Delta} &- a^2\sin^2\theta\Big]\frac{\partial^2\psi_0}{\partial t^2} + \frac{4Mar}{\Delta} \frac{\partial^2 \psi_0}{\partial t\partial\phi} + \Big[\frac{a^2}{\Delta} - \frac{1}{\sin^2\theta}\Big]\frac{\partial^2\psi_0}{\partial\phi^2}\\
        &-\Delta^{-2}\frac{\partial}{\partial r}\Big(\Delta^3 \frac{\partial\psi_0}{\partial r}\Big) - \frac{1}{\sin\theta}\frac{\partial}{\partial\theta}\Big(\sin\theta\frac{\partial}{\partial\theta}\Big) - 4\Big[\frac{a(r-M)}{\Delta} + \frac{i\cos\theta}{\sin^2\theta}\Big]\frac{\partial\psi_0}{\partial \phi}\\
        &-4\Big[\frac{M(r^2-a^2)}{\Delta} - r - ia\cos\theta\Big]\frac{\partial\psi_0}{\partial t} + (4\cot^2\theta - 2)\psi_0 = 0
    \end{split}
\end{equation}
\end{widetext}

Using a separation of variables $\psi_0 \sim e^{-i\omega t}e^{im\phi}S(\theta)R(r)$ it can be shown \cite{Teukolsky:1973ha},
\begin{widetext}

\begin{equation}
    \Delta^{-2}\frac{d}{dr}\Big(\Delta^3 \frac{dR}{dr}\Big) + \Big(\frac{K^2 - 4i(r-M)K}{\Delta} + 8i\omega r - \Tilde{\lambda}\Big)R = 0
\end{equation}
\begin{equation}
    \frac{1}{\sin\theta}\frac{d}{d\theta}\Big(\sin\theta\frac{dS}{d\theta}\Big) + \Big(a^2\omega^2\cos^2\theta - \frac{m^2}{\sin^2\theta} - 4a\omega \cos\theta - \frac{4m\cos\theta}{\sin^2\theta} - 4\cot^2\theta + 2 + A\Big)S = 0,
\end{equation}
\end{widetext}

where $K\equiv (r^2 + a^2)\omega - am$ and $\Tilde{\lambda} \equiv A + a^2\omega^1 - 2am\omega$. These sets of equations have been studied in details in the literature.

In our case, as we focus mainly on stationary perturbation, we will discuss such a scenario here. In that case, the solution can be represented as,

\begin{equation}
\psi_0 = \sum_{m=-2}^2\,  _2Y_{2m}(\theta,\phi)R_m(r).
\end{equation}

The solution of the $R_m(r)$ is of our main concern in this work. We expect it to satisfy the reflective boundary condition near the reflective surface. It can be formed from a linear combination of the linearly independent solutions of $R_m$. This has been found in Ref.~\cite{Teukolsky_thesis}. The two solutions are,

\begin{eqnarray}
y_1 =& x^{\gamma_m-2}(1+x)^{-\gamma_m-2} F(-l-2, l-1; -1+2\gamma_m; -x),\nonumber\\
y_2 =& x^{-\gamma_m}(1+x)^{\gamma_m} F(-l+2, l+3; 3-2\gamma_m; -x),
\end{eqnarray}

where $x=(r-r_+)/(r_+-r_-)$ and $\gamma_m = iam/(r_+-r_-)$ and $F$ is the hypergeometric function. Using these two, the relevant solution in Eq.(\ref{R_m}) has been found.

\bibliography{QBH_Heating.bib}

\end{document}